\documentclass[fleqn,usenatbib]{mnras}

\usepackage{newtxtext,newtxmath}
\usepackage{natbib}
\usepackage{stfloats}
\usepackage{lipsum}  
\usepackage{commath}

\usepackage[T1]{fontenc}

\DeclareRobustCommand{\VAN}[3]{#2}
\let\VANthebibliography\thebibliography
\def\thebibliography{\DeclareRobustCommand{\VAN}[3]{##3}\VANthebibliography}

\usepackage{graphicx}	
\usepackage{amsmath}	
\usepackage[list=no]{subcaption}
\usepackage{float}
\usepackage{xcolor}

\usepackage{comment} 

\usepackage{amsmath}

\title[Stellar bar occurrence: disentangling environmental effects and assembly history]{Stellar bar occurrence in TNG50 and TNG100: disentangling environmental effects and assembly history at z = 0.}

\author[Chim Ramirez]{
Karol Chim-Ramirez,$^{1}$\thanks{E-mail: k.chim@irya.unam.mx} 
Bernardo Cervantes Sodi,$^{1}$
Yetli Rosas-Guevara,$^{2}$
\\
$^{1}$Instituto de Radioastronomía y Astrofísica, Universidad Nacional Autónoma de México, Antigua Carretera a Pátzcuaro 8701, Ex-Hda. San José \\ de la Huerta, Morelia, Michoacán, México C.P. 58089\\
$^{2}$Departamento de Física, Universidad de Córdoba, Campus Universitario de Rabanales, Ctra. N-IV Km. 396, E-14071 Córdoba, Spain
}

\date{Accepted XXX. Received YYY; in original form ZZZ}

\pubyear{2026}

\usepackage{xcolor} 

\begin{document}
\label{firstpage}
\pagerange{\pageref{firstpage}--\pageref{lastpage}}
\maketitle

\begin{abstract}
In this work, we investigate the dependence of bar presence on environment using TNG50 and TNG100 simulations at $z=0$. We select disc galaxies within $2 R_{200}$, with stellar mass M$_{\star} \geq 10^{10}$ M$_{\odot}$ (TNG100) and M$_{\star} \geq 10^{9}$ M$_{\odot}$ (TNG50). With these criteria, our final samples consist of $1719$ and $859$ satellite galaxies in TNG100 and TNG50, respectively. The environment is characterised by host halo mass: massive (M$_{\rm h} \geq 10^{14}$ M$_{\odot}$), intermediate ($10^{13} < \rm M_{\rm h} < 10^{14}$ M$_{\odot}$), and low-mass haloes (M$_{\rm h} \leq 10^{13}$ M$_{\odot}$). We find that our samples recover the morphology–density relation, with thin disc fractions increasing with cluster-centric distance and decreasing halo mass. The bar fraction and the bar radius normalized by the effective radius decrease mildly with cluster-centric distance, while absolute bar radii remain constant. At fixed distance, the bar fraction increases with halo mass, partly because more massive galaxies reside in more massive haloes. In addition, we find that galaxies with earlier assembly times show a higher bar fraction, especially within $0.5 R_{200}$, suggesting a connection between earlier assembly and a higher probability of hosting a bar. Overall, we find a mild environmental dependence, with a stronger connection to stellar mass and assembly history.

\end{abstract}




\section{Introduction}
Stellar bars are galaxy features present in around $30 \% - 50 \%$ of disc galaxies in the universe \citep{sellwood1993dynamics, eskridge2000frequency, masters2011galaxy}. They are non-axisymmetric structures located in the inner regions of their host galaxies. Bars play an important role in the secular evolution of galaxies \citep{kormendy2013secular} by redistributing mass and angular momentum among the different galaxy components \citep{long2014secular, sellwood1980galaxy, holley2005bar}.
This evolution is governed by the interplay between the bar, stellar disc, and dark matter halo. In particular, dynamical friction exerted by the halo can slow down the rotation of the bar \citep{weinberg1985evolution}, as shown by N-body simulations \citep{debattista1998dynamical, debattista2000constraints}, while resonant interactions with the halo can also promote bar growth \citep{athanassoula2002formation}. Bar formation is a complex process that depends on several physical parameters. In general, it requires a sufficiently massive, dynamically cold disc \citep{ostriker1973numerical, efstathiou1982stability} with low velocity dispersion and predominantly rotational support \citep{kalnajs1978proc, toomre1981amplifies, christodoulou1995}, allowing disc instabilities to develop and give rise to the formation of the bar, as also confirmed in cosmological simulations \citep{algorry2017barred, Romeo2023specific}. These conditions not only favor bar formation but also its subsequent growth, during which gas and stars can be driven toward the central regions of the galaxy, contributing to pseudobulge formation and enhanced central star formation. This secular evolution has been discussed extensively in the review by \citet{kormendy2004secular} and supported by observational studies \citep{drory2007connection, fisher2008structure, fisher2010bulges}. Numerical simulations have likewise shown that bars can drive the redistribution of gas and stars toward the central regions, promoting the growth of pseudobulges and central mass concentrations \citep{athanassoula2005nature, athanassoula2013bars, debattista2006secular, saha2013secular, khoperskov2018bar}.

Bar formation and evolution could be affected by several galactic parameters. Several studies, both observational \citep{diaz2016characterization, sodi2017low} and theoretical \citep{algorry2017barred, rosas2020buildup, zhao2020barred, chim2025study} have found that the bar fraction, that is the ratio between the number of barred galaxies and the total number of disc galaxies in the sample, is strongly correlated with stellar mass, such that the bar fraction increases with increasing stellar mass. This behaviour is consistent with the early formation of bars in massive galaxies \citep{vera2016effect}. The bar instabilities in the disc take place earlier in these galaxies and occur faster, with a timescale of 1 Gyr due to their stronger gravity and denser discs \citep{friedli1993secular, martel2013connection}. Other internal parameters that affect the presence of bars in galaxies are the amount of gas \citep{friedli1993secular, combes2009secular, athanassoula2013bar, seo2019effects}, the presence (or not) of a classical bulge \citep{chacon2024, rosas2025Galaxy, chim2025study, kwak2026smuggle}, the dark matter halo fraction \citep{sodi2015dark, cervantes2017stellar} and spin \citep{cervantes2013galactic, aquino2026bar}. 

In addition to the internal parameters, the environment where galaxies reside could also play a role in galaxy formation and evolution \citep{moore1998morphological, mastropietro2005morphological, cerulo2016accelerated, cerulo2017morphological}. When a galaxy is located in a populated environment, phenomena like tidal interactions, mergers, ram pressure stripping, and others could cause a morphological transformation \citep{dressler1980galaxy, vulcani2023clustercentric}. A key observational manifestation of this is the morphology-density relation, which shows that early-type galaxies (ellipticals and lenticulars) preferentially inhabit high-density regions like clusters, while late-type galaxies (spirals) are more common in low-density environments \citep{dressler1980galaxy}. Currently, it is not known exactly how these interactions and the environment can affect stellar bars \citep{lokas2016tidally, zana2018external}, which arise from disc perturbations. Theoretical and observational studies present varied and sometimes contradictory conclusions. Understanding the role of environment is essential, as it provides insight into the physical processes that regulate disc instabilities and, consequently, the secular evolution of galaxies. In particular, environmental effects with the cluster potential can significantly alter the dynamical state of galactic discs \citep{Ding2024Effects, Mendez2023JWST}.

On the one hand, the majority of observational studies seem to point in the same direction: the formation of stellar bars appears to be inhibited in galaxies experiencing some form of interaction with a near galaxy. It has also been found that strong bars can be destroyed when galaxies undergo intense tidal interactions within the inner regions of a galaxy cluster \citep{lee2012dependence, lin2014environment, tawfeek2022morphology}. Despite this general agreement among most observational works, some studies have reported conflicting results. For instance, \cite{thompson1981bar} and \cite{andersen1996distribution} discovered that, in the case of the Coma and Virgo clusters, respectively, the bar fraction tends to increase in proximity to the cluster centre. This can be attributed to the induction of disc instabilities by strong interactions, which subsequently result in the formation of bars in galaxies that were initially stable. However, other more recent works (e.g. \citealt{mendez2010galaxies, lansbury2014barred}) concluded that the bar fraction does not vary significantly when moving from the centre to the periphery of the cluster and that multiple interactions would heat the disc, preventing the formation of bars. In \cite{tawfeek2022morphology}, barred galaxies within a cluster environment taken from the OmegaWINGS survey were examined, and the authors found that the fraction of barred galaxies tends to decrease when considering increasingly inner regions of the clusters, a trend that vanishes once the morphological type is taken into account, indicating that the likelihood of a galaxy hosting a bar is determined by the morphological transformation triggered by the environment. The differences between these various works may arise from differences in sample selection, bar detection methods, and the range of environments probed, making it difficult to establish a unified picture.

Regarding theoretical studies using simulations, the issue seems to be less clear, as different works point to different conclusions. Some studies indicate that interactions do not play a significant role in bar formation and that, regardless of them, the bar will inevitably form with a sufficiently massive disc that promotes bar instabilities. In fact, in cosmological zoom-in simulations, interactions such as mergers or flybys could even delay the formation time of bars if such a structure is present, although once the bar is formed, they are not factors that affect the subsequent evolution of its properties (\citealt{zana2018external}). Other studies of N-body simulations (i.e. \citealt{lokas2016tidally, lokas2020tidal, smith2021brought}) conclude that tidal interactions can induce bar formation in the innermost parts of clusters and that these bars turn out to be stronger, longer, and formed earlier. According to \cite{pettitt2018bars}, which use N-body simulations, interactions slightly accelerate the formation of the bar when the galaxy rotation curve is dominated by the disc, unlike isolated galaxies or those with weak interaction where dark matter dominates. Thus, interactions seem to favor the formation and growth in these studies. \cite{rosas2024rise}, using the TNG50 magneto-hydrodynamical simulation, found that about one third of disc galaxies hosting a bar at ($z=1$) retain it until ($z=0$). Surviving bars tend to become stronger and longer with time, while differences in the merger histories of galaxies that retain or lose their bars suggest that interactions can influence bar survival without necessarily being the primary mechanism driving bar formation.

As we have seen, the literature presents diverse and sometimes conflicting results regarding how the environment in which galaxies reside affects their structural components, such as stellar bars and overall morphology. This motivates a systematic study of environmental effects within a consistent framework, enabling us to disentangle the role of different environmental mechanisms. In this work, we explore how the environment, characterised by the mass of the host cluster and the galaxy cluster-centric distance, impacts the presence of stellar bars at $z=0$. We consider different ranges of host halo mass and galaxy positions within the cluster to disentangle environmental effects. In addition, we analyze the morphological properties of galaxies using a morphological classification based on stellar kinematics, as well as the properties of bars, including their length and strength. We also investigate the assembly time of galaxies to explore possible connections between their formation history and the presence of bars. To achieve this, we employ the cosmological simulations from the IllustrisTNG project, specifically the highest-resolution simulation (TNG50) and the intermediate-resolution one (TNG100).

This paper is organized as follows. In Section \ref{methodology}, we describe the simulation, sample construction, and methods used in our analysis. In Section \ref{results}, we present our main results on the dependence of bar presence and properties on the environment. Finally, in Section \ref{discussion}, we discuss our findings and summarize our main conclusions.

\section{Methodology}
\label{methodology}

\subsection{IllustrisTNG Simulation}

The IllustrisTNG project (The Next Generation Illustris; \citealt{nelson2018first, pillepich2018first, naiman2018first, marinacci2018first, springel2018first}) is a suite of large-scale cosmological magneto-hydrodynamical simulations that follow the formation and evolution of galaxies within a $\Lambda$CDM framework consistent with the Planck Collaboration results (\citealt{Plack2016planck}). The simulations adopt cosmological parameters of $\Omega_b = 0.0486$, $\Omega_\Lambda = 0.6911$, $\Omega_m = 0.3089$ and $H_0 = 67.74$ km s$^{-1}$ Mpc$^{-1}$. 

Dark matter haloes in IllustrisTNG are identified using a Friends-of-Friends (FoF) algorithm with a linking length of $l = 0.2$ times the mean particle separation \citep{davis1985evolution}. Within these FoF groups, gravitationally bound substructures —interpreted as galaxies— are located and hierarchically classified. Subsequently, the \texttt{SUBFIND} algorithm \citep{springel2001populating} is applied to detect gravitationally bound substructures, providing the identification of individual galaxies within the FoF haloes \citep{pillepich2018simulating}. In the simulation, the central galaxy is typically defined as the most massive substructure located near the centre of the FoF halo, while the remaining substructures are classified as satellite galaxies. Each FoF halo therefore contains one central galaxy and, in most cases, multiple satellites.

In this paper, we employ the TNG50 and TNG100 simulations. Both runs contain $100$ snapshots spanning from $z=20$ to $z=0$. For the present work, we use snapshot 99, corresponding to redshift $z=0$. For our analysis, we focus on galaxies with stellar masses above $10^9$ M$_{\odot}$ in TNG50 and $10^{10}$ M$_{\odot}$ in TNG100. These thresholds ensure well-resolved galaxies consistent with the resolution limits of each simulation. The resulting parent samples comprise $3295$ and $6507$ galaxies (centrals and satellites) in TNG50 and TNG100, respectively. Given that the objective of this work is to analyze how the bar presence is affected by the environment and galaxy evolution, we also characterise the formation histories through the stellar mass assembly time, $z_{0.5 M_{\star}}$, defined as the redshift at which each galaxy assembled half of its final stellar mass.

\subsection{\texttt{MORDOR} algorithm}

In order to identify the morphological components of galaxies in our samples, we used the \texttt{MORDOR} (MORphological DecOmposeR) algorithm (\citealt{zana2022mordor}). This code splits the stellar particles according to two parameters: their energy, defined by $E= 1/2 |\mathrm{v}|^2 + \psi$, where $\mathrm{v}$ is the particle velocity and $\psi$ is the gravitational potential, and their circularity $\eta \equiv j_z/j_{circ}(E)$. Once these parameters have been calculated, the algorithm splits the energy distribution according to the lowest minimum, called $E_{cut}$. With this, \texttt{MORDOR} segregates the particles into the most and less bound particles and identifies five different components: thin and thick disc, bulge, pseudobulge, and stellar halo. Here we provide only a brief summary of the algorithm; for a detailed description, we refer the reader to \cite{zana2022mordor}.

Three morphological components are identified among the most bound particles. Stellar particles with circularities $\eta<0$ are assigned to the bulge, and an equivalent distribution with $\eta>0$ is added through Monte-Carlo sampling to enforce symmetry about $\eta=0$. Of the remaining particles, those with $\eta>0.7$ are classified as the thin disc, while those with $0\leq\eta\leq0.7$ are assigned to the pseudobulge. For the less bound particles, the same procedure is used to identify the stellar halo: particles with $\eta<0$ are assigned to the halo, together with a Monte Carlo-sampled distribution with $\eta>0$ that ensures symmetry in the population. The remaining particles are then divided into the thin disc ($\eta>0.7$) and thick disc ($0\leq\eta\leq0.7$).  We restrict our analysis to galaxies resolved with $10^4$ stellar particles or more, corresponding to stellar masses above $10^9$ M$_{\odot}$ in TNG50 and $10^{10}$ M$_{\odot}$ in TNG100. These limits ensure that the morphological decomposition is well resolved. 

\begin{figure*}
\captionsetup[subfigure]{labelformat=empty}
\begin{subfigure}{0.81\linewidth}
    \includegraphics[width=\linewidth, height=0.4\linewidth]{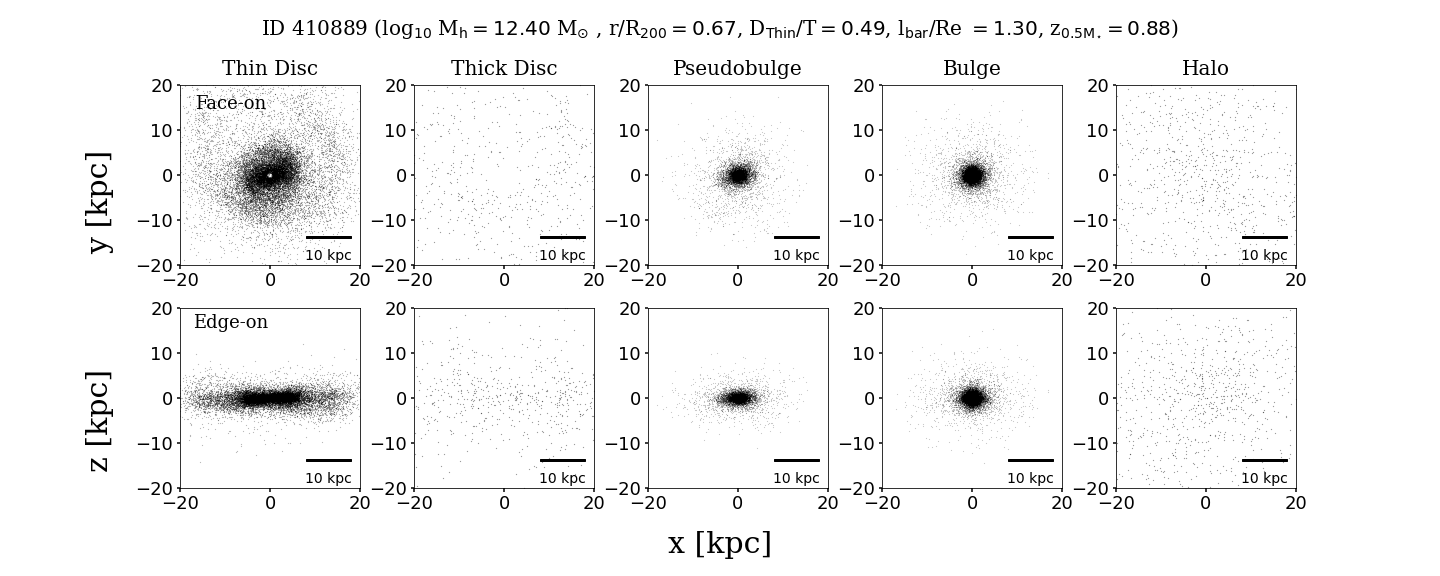}
    \subcaption{ }
    \label{mordorpar}
\end{subfigure}
    \hfill
\begin{subfigure}{0.81\linewidth}
    \includegraphics[width=\linewidth, height=0.4\linewidth]{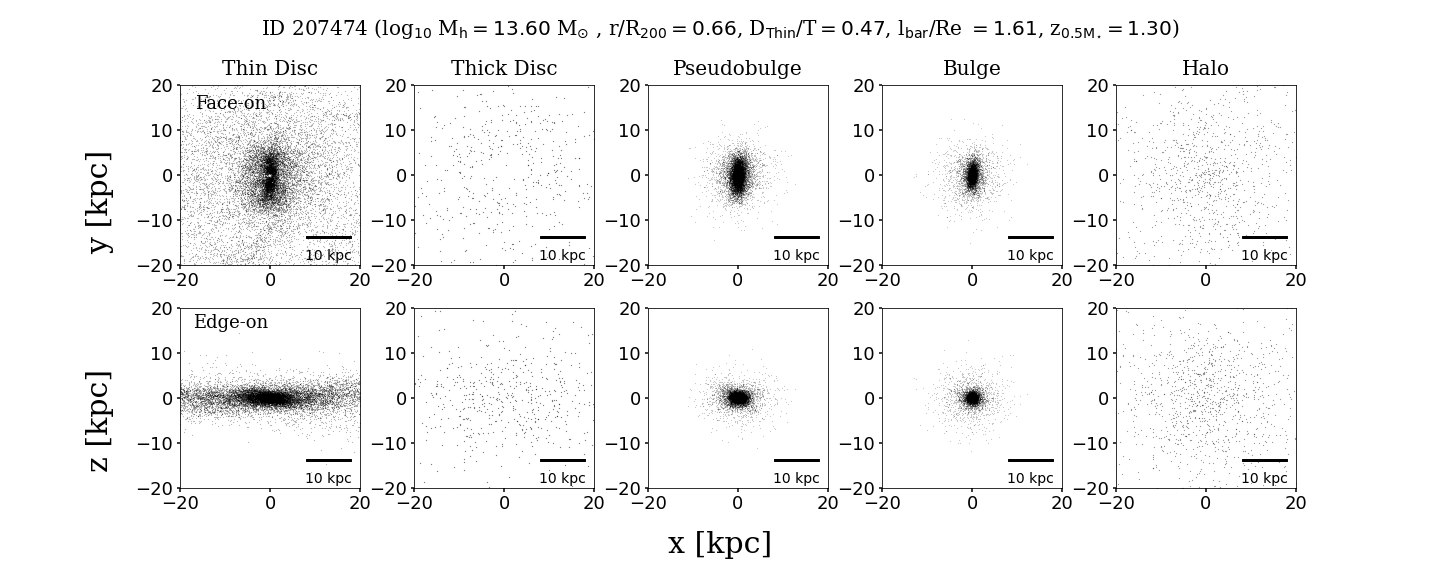} 
  \subcaption{ }
  \label{mordorgroup}
\end{subfigure}
    \hfill
\begin{subfigure}{0.81\linewidth}
    \includegraphics[width=\linewidth, height=0.4\linewidth]{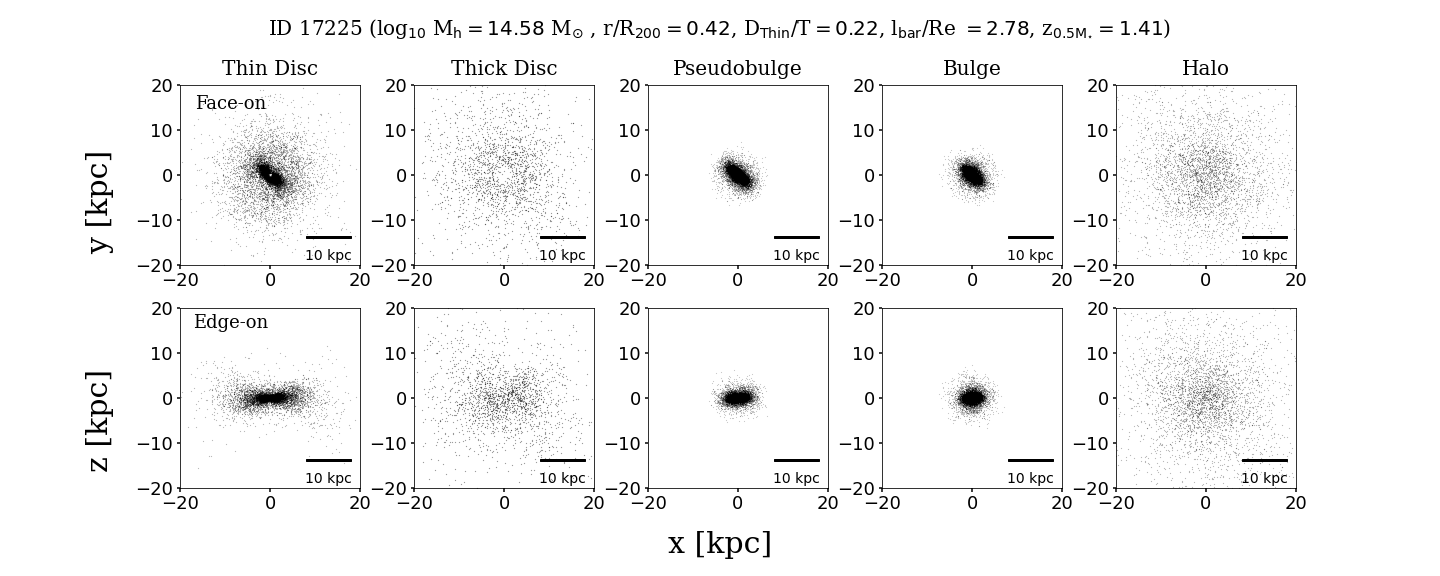} 
  \subcaption{ }
  \label{mordorcluster}
\end{subfigure}
\caption{Morphological components identified by the \texttt{MORDOR} algorithm (\citealt{zana2022mordor}) for three TNG100 galaxies with similar stellar mass ($\log_{10} \mathrm{M}_{\star} \sim 10.5 \mathrm{M_{\odot}}$) but residing in distinct environments within the virial radius. The upper panel shows a galaxy in a low-mass halo at $0.67$ R$_{200}$ from the cluster centre. The middle panel corresponds to a galaxy in an intermediate-mass halo with a cluster-centric distance of $\sim 0.66 R_{\rm 200}$. Finally, the lower panel corresponds to a galaxy in one of the most massive clusters at $0.42 R_{\rm 200}$. We appreciate that, across all panels, the stellar bar structure is clearly visible in the inner regions of the galaxies, regardless of halo mass, and that all lie near the cluster centres. In each panel, the thin disc component, normalized bar size, and assembly time are indicated at the top. The marker size for the thick disc and halo components has been increased to improve their visibility, as these components contain fewer particles than the others. Each panel includes a scale bar indicating the physical size.}
\label{mordorgalaxies}
\end{figure*}

For the TNG50 galaxies, we used the catalogue provided by \citet{zana2022mordor}. For the TNG100 sample, we ran \texttt{MORDOR} on all galaxies in our sample. However, in some galaxies in TNG100, particles with high circularity ($\eta \gtrsim 0.7$), which are kinematically characteristic of discs, were being assigned to the stellar halo. To prevent this contamination, we introduced a modification to the algorithm by imposing a circularity threshold of $\eta \leq 0.7$ for particles that can be considered for the halo. This ensures that the halo is populated predominantly by particles with low circularity, while high-circularity particles are correctly assigned to the thin or thick disc. This modification preserves the symmetry principle of the original method while avoiding the artificial mixing of disc and halo populations that can arise in simulations with lower particle resolution. In Fig. \ref{mordorgalaxies}, we show the particle distribution of each morphological component for three barred satellite galaxies in TNG100 with similar stellar masses (M$_{\star} \sim 10^{10.5}$ M$_{\odot}$), each residing in a different environment (discussed in detail in Section \ref{enviroment_charact}). At the top of each panel, we indicate the galaxy ID, the logarithm of mass of its host halo ($\log_{10}$ M$_{\rm h}$), the normalized distance to the central galaxy of the halo ($r/R_{200}$), the thin disc component (D$_{\rm Thin}$/T), the normalized bar radius (l$_{\rm bar}$/Re), and the stellar mass assembly redshift (z$_{0.5 \rm M_{\star}}$). The upper panel shows a galaxy located in a low-mass halo ($\log_{10}$ M$_{\rm h}$ $= 12.40$ M$_{\odot}$) at $0.67$ R$_{200}$; it exhibits a prominent bar in the face-on view, and spiral arms are also visible. The middle panel displays a galaxy in an intermediate-mass halo ($\log_{10}$ M$_{\rm h}$ $= 13.60$ M$_{\odot}$), located at $0.66 R_{200}$; this galaxy also hosts a bar. Finally, the lower panel shows a galaxy in a massive halo ($\log_{10}$ M$_{\rm h}$ $= 14.58$ M$_{\odot}$), relatively close to the cluster centre with $r/R_{200} = 0.42$. As shown in the panels, the \texttt{MORDOR} algorithm assigns part of the bar particles to the bulge and pseudobulge components, rather than to the disc, because bar stars have lower angular momentum than stars on nearly circular orbits. We therefore deliberately use the thin disc fraction as a conservative indicator of the cold disc component, rather than as an estimate of the total disc mass, particularly in barred galaxies.

\subsection{Bar identification}
To identify stellar bars, we apply a Fourier decomposition to the face-on stellar surface density (\citealt{athanassoula2002morphology, rosas2020buildup}). This bar identification is performed independently of the \texttt{MORDOR} kinematic decomposition, meaning that the morphological classification of a galaxy into components does not influence the bar detection. To identify a bar structure, we focus on two parameters, the  ratio between the second and zero terms of the Fourier expansion $A_2(R)$ and the phase $\Phi(R)$:

\begin{equation}
    A_2 (R)= \dfrac{|\sum_j m_j e^{2i \theta_j}|}{\sum_j m_j}
\end{equation}

\begin{equation}
    \Phi(R) = \dfrac{1}{2} \arctan \left[ \dfrac{\sum_j m_j \sin{2 \theta_j}}{\sum_j m_j \cos{2 \theta_j}} \right] ,
\end{equation}

where $m_j$ is the mass of the j-th stellar particle and $\theta_j$ is its angular position. This calculation is performed within twice the stellar half-mass radius, considering stellar particles located within $\pm 1$ kpc along the $z$-axis and distributed in coaxial cylindrical shells of width $0.12$ kpc. The bar strength is defined as the maximum value of $A_2$, while the bar radius l$_{\rm bar}$ is identified as the outermost radius where the phase angle $\phi$ remains approximately constant, indicating a coherent non-axisymmetric structure. Specifically, we search for contiguous radial regions where the total phase variation does not exceed $8^{\circ}$ and require that valid bar regions have a minimum radial extent of 1 kpc, a commonly adopted criterion in several studies (e.g. \citealt{rosas2022evolution, zana2022mordor}) to identify the bar region, where the phase variation is typically required to be smaller than $0.1$  –$0.15$ radians ($\sim 6$–$8^{\circ}$). In both simulations, a galaxy is classified as barred if it exhibits a bar strength $A_2$ greater than $0.2$ and a minimum bar radius of $1$ kpc. To avoid spurious detections, the bar identification method was applied only to galaxies with dominant rotational support (i.e. disc galaxies), quantified by $k_{\mathrm{rot}} \geq 0.4$ (\citealt{sales2010feedback}). This parameter was computed as

\begin{equation}
\centering
k_{\mathrm{rot}}= \frac{K_{\mathrm{rot}}}{K}= \frac{1}{K} \sum \frac{1}{2} m_i \left( \frac{j_{z,i}}{R_i} \right)^2 ,
\end{equation}

where $K_{\mathrm{rot}}$ corresponds to the kinetic energy of the stellar component along the azimuthal direction, and $K$ is the total kinetic energy. The sum is over all stellar particles in the galaxy, where $m_i$ is the particle mass, $j_{z,i}$ is the specific angular momentum along the $z$-axis, and $R_i$ is the projected radius. It is important to emphasise that $k_{\mathrm{rot}}$ measures the overall rotational support of the stellar component and does not distinguish between co-rotating and counter-rotating material. Nevertheless, it remains a robust indicator of rotational support, as significant counter-rotating stellar discs are rare: only $\sim 10 \%$  of Milky Way-mass late-type galaxies in TNG50 host one \citep{burgeno2025origin}, consistent with observational estimates of $\sim 1 \%$  in late-type galaxies \citep{bao2022counter, gasymov2025counter}. Applying this criterion reduces our samples to $4953$ and $2639$ galaxies for TNG100 and TNG50, respectively, including both centrals and satellites. From these, we identified $2074$ barred galaxies in TNG100 and $517$ in TNG50.

\begin{figure*}
\captionsetup[subfigure]{labelformat=empty}

\begin{subfigure}{\linewidth}
    \includegraphics[width=\linewidth]{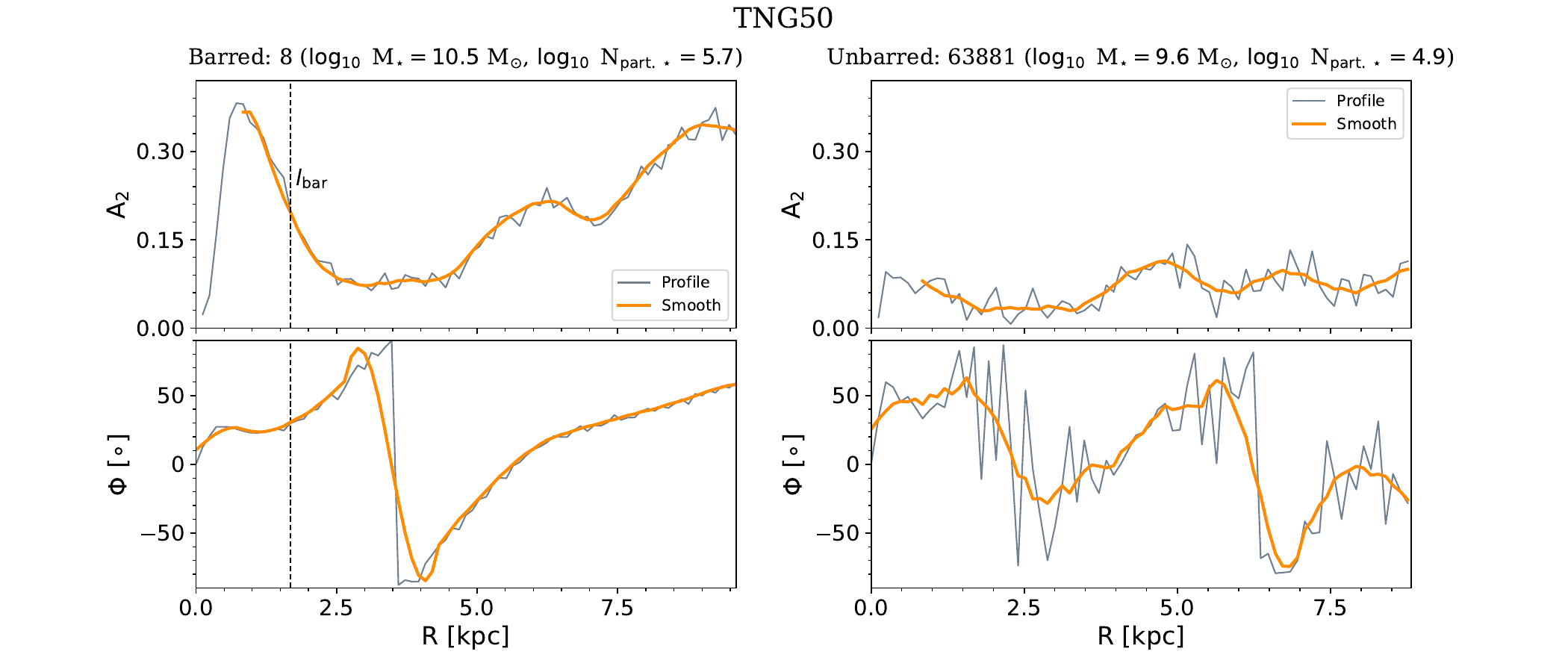}
    \subcaption{}
    \label{TNG50bFouier}
\end{subfigure}

\vspace{0.4cm}

\begin{subfigure}{\linewidth}
    \includegraphics[width=\linewidth]{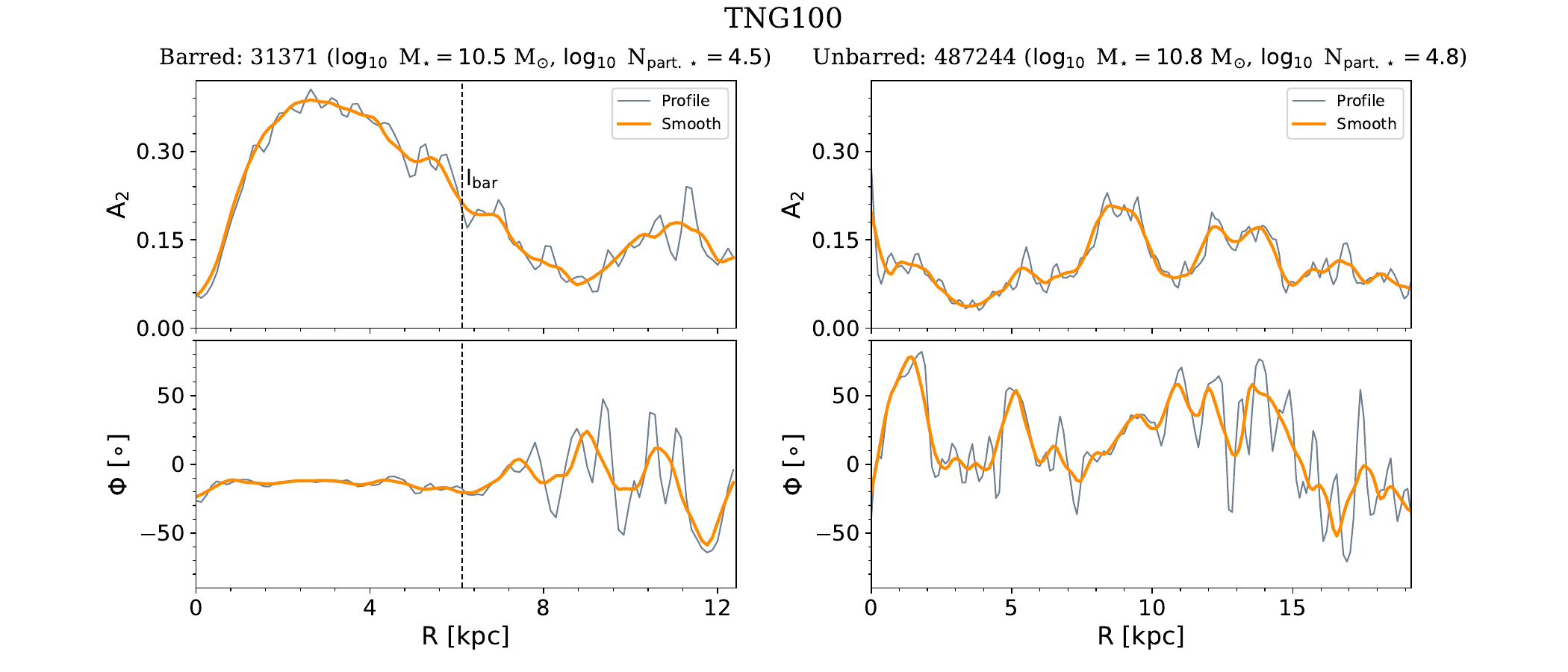 }
    \subcaption{}
    \label{TNG100Fourier}
\end{subfigure}

\caption{A$_2$ and phase $\Phi$ profiles as a function of radius $R$ for barred (left) and unbarred (right) galaxies randomly selected from TNG50 and TNG100, shown in the upper and lower panels, respectively. We observe that barred galaxies exhibit an increase in A$_2$ in the inner regions and a nearly constant phase $\Phi$, while unbarred galaxies do not show a clear A$_2$ peak and display highly variable phase profiles. The black dotted vertical lines in barred galaxies indicate the bar radius. The galaxy ID, stellar mass, and number of stellar particles are indicated at the top of each panel.}
\label{FourierTNG50and100}
\end{figure*}

In Fig. \ref{FourierTNG50and100}, we show the $A_2$ and phase $\Phi$ radial profiles for both simulations, TNG50 (upper panels) and TNG100 (lower panels), separating barred (left) and unbarred (right) systems. As expected, barred galaxies display a pronounced increase of $A_2$ in their central regions and a nearly constant phase $\Phi$ across the bar extent, whereas unbarred galaxies show no significant $A_2$ peak and exhibit strong phase variations. The dotted lines indicate the estimated bar radii. These examples illustrate the expected profiles for barred and unbarred galaxies.

Since $k_{\mathrm{rot}}$ serves as a proxy for morphology, we verified that our selection is consistent with late-type galaxies, finding good agreement with morphological decompositions. 

\subsection{Environment characterisation.}
\label{enviroment_charact}
To analyse how galaxy and bar properties depend on environment, we use two main parameters: the halo mass, M$_{\rm h}$, in which galaxies reside, and the cluster-centric distance $r/R_{200}$. Here, M$_{\rm h}$ is defined as M$_{200}$, the mass enclosed within the radius $R_{200}$, where $R_{200}$ is the radius within which the mean density of the halo is 200 times the critical density of the Universe. The cluster-centric distance $r/R_{200}$ is defined as the ratio between the distance from a galaxy to the central galaxy and the $R_{200}$ of its respective halo. 
The main focus of this study is the satellite disc galaxies with a cluster-centric distance $r/R_{200}$ smaller than 2; this limit is motivated by the distribution of $r/R_{200}$ in our parent samples. With this restriction, our samples of satellite galaxies are reduced to $1719$ and $859$ galaxies for TNG100 and TNG50, respectively, while maintaining the central galaxies subsample for comparison in the following section.

\begin{figure*}
\captionsetup[subfigure]{labelformat=empty}

\begin{subfigure}{0.5\linewidth}
    \centering
    \includegraphics[width=\linewidth]{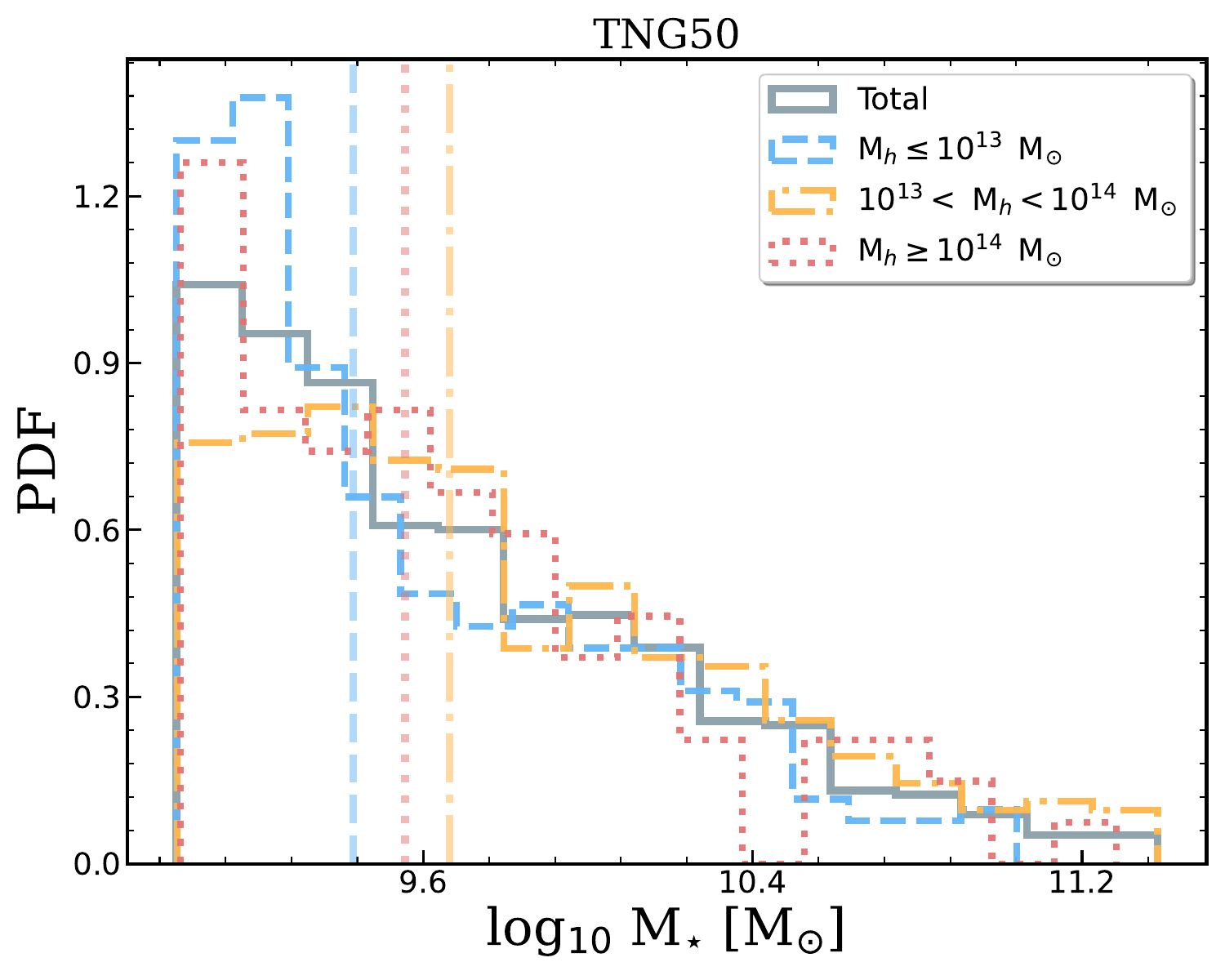}
    \subcaption{}
    \label{Ms_halo_TNG50}
\end{subfigure}%
\hspace{-0.01\linewidth}
\begin{subfigure}{0.5\linewidth}
    \centering
    \includegraphics[width=\linewidth]{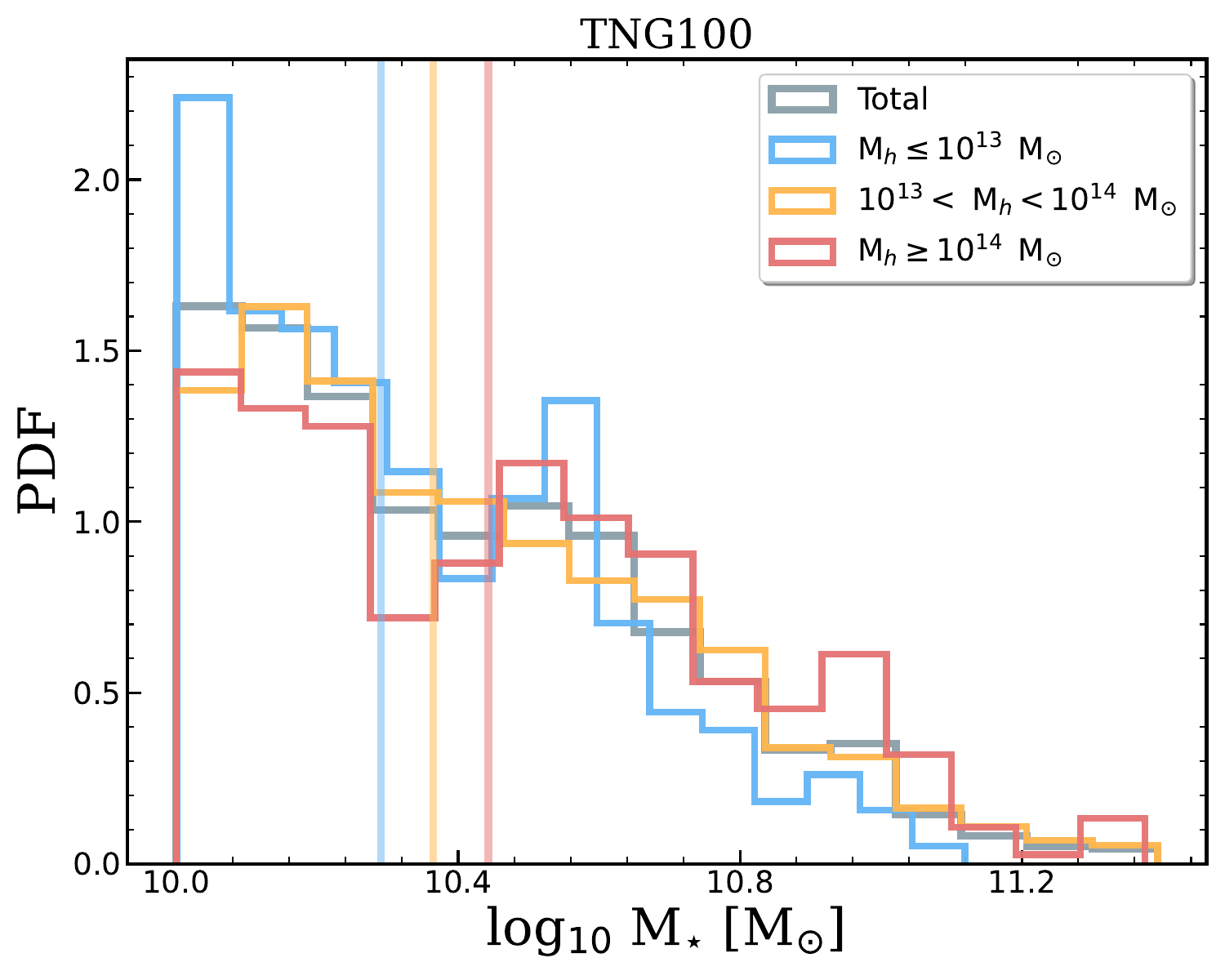}
    \subcaption{}
    \label{Ms_halo_TNG100}
\end{subfigure}

\caption{Probability density function (PDF) of stellar mass for galaxies in the different halo mass ranges, shown for TNG50 (left panel) and TNG100 (right panel). The different colours and line styles correspond to the halo mass ranges indicated in the legend. Vertical lines mark the median stellar mass for each halo mass range. } 
\label{dist_Ms_haloes}
\end{figure*}

To characterise the environments where our galaxies reside, we classify haloes according to their halo mass: massive haloes (M$_h \geq 10^{14}$ M$_{\odot}$), intermediate-mass haloes ($10^{13}$ < M$_h < 10^{14}$ M$_{\odot}$), and low-mass haloes (M$_h \leq 10^{13}$ M$_{\odot}$). Each halo consists of one central galaxy (the most massive member) and several satellites. Table \ref{numhaloes} lists the number of haloes found in each simulation, and for each halo mass range, the number of galaxies and bars. As expected, TNG50 contains only one massive halo (and generally fewer haloes across all mass ranges) compared to TNG100, which covers a larger volume. In Fig. \ref{dist_Ms_haloes}, we show the stellar mass distributions of galaxies within different halo mass ranges. As can be seen, the distributions for both total samples are similar, with the majority of galaxies concentrated at lower stellar masses. This trend is also observed for low- and intermediate-mass haloes in both simulations. In contrast, galaxies in the most massive TNG100 haloes exhibit an almost homogeneous distribution across the stellar mass range.

\begin{table*}
\centering
\setlength{\tabcolsep}{12pt} 
\caption{\textbf{Number of haloes (N$_{\mathrm{haloes}}$), total galaxies (N$_{\mathrm{gal}}$), and barred galaxies (N$_{\mathrm{bars}}$) in different halo mass range for the TNG100 and TNG50 simulations.}}
\label{numhaloes}
\begin{tabular}{lccccccccc}
\hline
 & \multicolumn{3}{c}{$\log_{10}(\mathrm{M}_{\mathrm{halo}}$ $/ \mathrm{M}_\odot) \leq 13$} 
 & \multicolumn{3}{c}{$13 < \log_{10}(\mathrm{M}_{\mathrm{halo}}/ \mathrm{M}_\odot) < 14$} 
 & \multicolumn{3}{c}{$\log_{10}(\mathrm{M}_{\mathrm{halo}}/ \mathrm{M}_\odot) \geq 14$} \\
\cline{2-4} \cline{5-7} \cline{8-10}
Simulation & N$_{\mathrm{haloes}}$ & N$_{\mathrm{gal}}$ & N$_{\mathrm{bars}}$ & $N_{\mathrm{haloes}}$ & $N_{\mathrm{gal}}$ & $N_{\mathrm{bars}}$ & $N_{\mathrm{haloes}}$ & $N_{\mathrm{gal}}$ & $N_{\mathrm{bars}}$ \\
\hline
\textbf{TNG100} & 391 & 515 & 213 & 160 & 794 & 373 & 14 & 410 & 230 \\
\textbf{TNG50}  & 226 & 379 & 96  & 23  & 391 & 123 & 1  & 89  & 32  \\
\hline
\end{tabular}
\end{table*}

\section{Results}
\label{results}
In this section, we present the results of our analysis for our disc galaxy samples.

\subsection{Morphology-density relation.}
One of the most important relations for galaxies is the connection between their morphology and the environment in which they reside. This is widely known as the morphology–density relation (\citealt{dressler1980galaxy}), which states that the morphology of galaxies depends on the local galaxy density. The fraction of early-type galaxies, such as ellipticals and lenticulars, increases with galaxy density, while the fraction of late-type galaxies, such as spirals, decreases with this parameter (\citealt{oemler1974systematic, dressler1980galaxy, postman1984morphology, fasano2015morphological}). This relation is also found in simulations (e.g., \citealt{okamoto2001morphology, lotz2019gone, joshi2020fate, pfeffer2023galaxy}).

\begin{figure*}
\captionsetup[subfigure]{labelformat=empty}
\centering
    \begin{subfigure}{0.5\linewidth}
        \includegraphics[width=\linewidth]{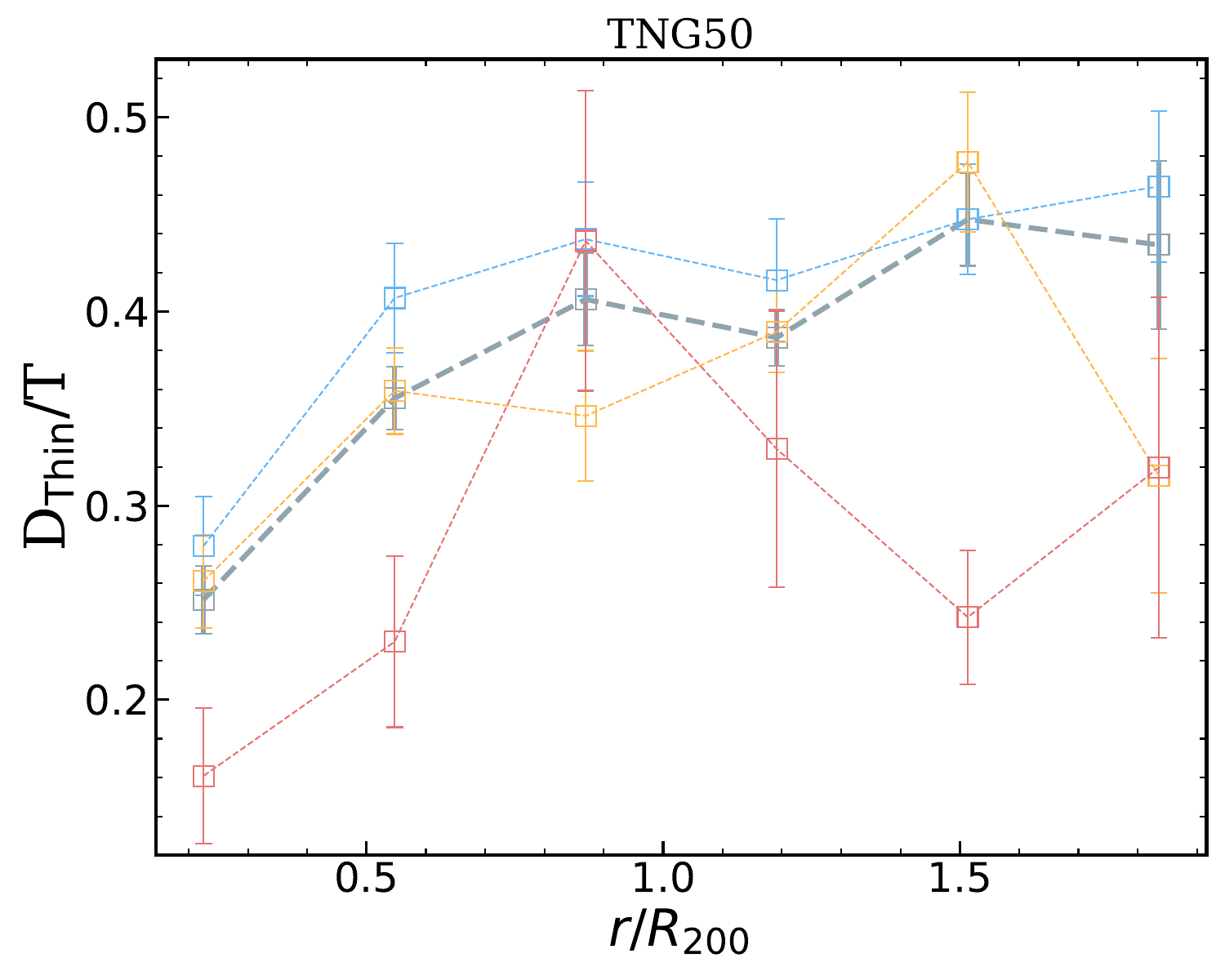}
        \subcaption{ }
        \label{fig:sfig1}
    \end{subfigure}
    \hspace{-0.01\linewidth}
    \begin{subfigure}{0.5\linewidth}
        \includegraphics[width=\linewidth]{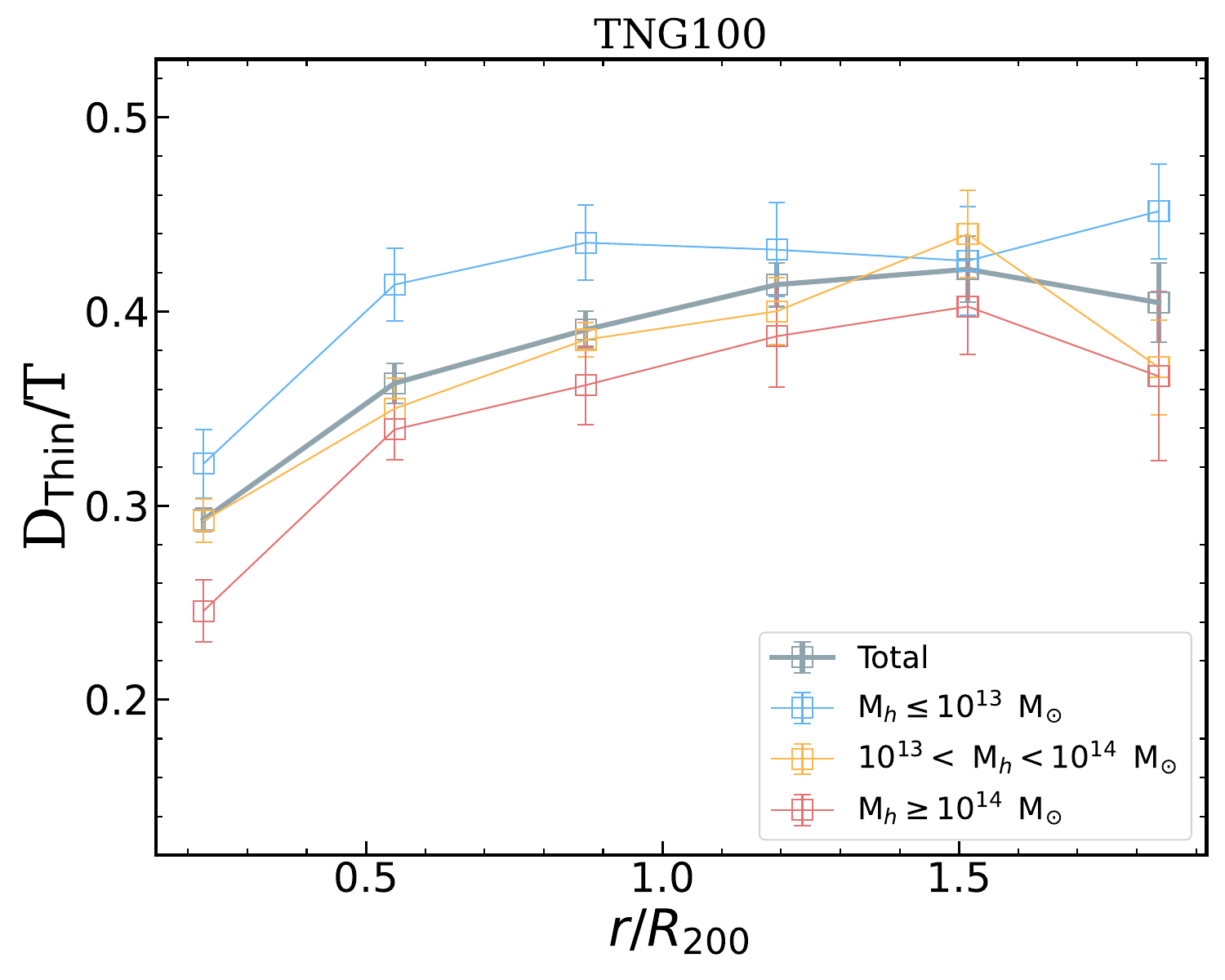}
        \subcaption{ }
        \label{fig:sfig2}
    \end{subfigure}

\caption{Thin disc component D$_{\rm Thin}$/T as a function of cluster-centric distance for TNG50 (left panel) and TNG100 (right panel), shown for the total samples (gray lines) and for low- (blue), intermediate- (yellow), and massive-halo (red) mass ranges. We observe that galaxies in the samples clearly reproduce the morphology–density relation, with higher thin disc components at larger distances from the cluster centre. We also find that galaxies in low-mass haloes exhibit higher thin disc components. Note that for TNG50, the massive-halo sample (log$_{10}$ M$_{\rm h} \geq 10^{14}$ M$_{\odot}$) contains only one host halo and therefore does not represent a population of clusters in this mass range.} 
\label{ThinDTNG50y100}
\end{figure*}

To analyze this relation in our subsamples, we use the thin disc component, D$_{\rm Thin}$/T, defined as the fraction of thin disc mass (obtained from \texttt{MORDOR}) with respect to the total stellar mass, as a proxy for morphology. In Fig. \ref{ThinDTNG50y100} we show the thin disc fraction as a function of the cluster-centric distance $r/R_{200}$ for different halo mass bins. In what follows, plots with dotted lines correspond to the TNG50 simulation, while solid lines correspond to TNG100. The error bars were calculated using the bootstrapping resampling method and a 1 $\sigma$ confidence level. All error bars in the following figures were calculated using this method. The different colours represent the different halo masses. As can be seen in Fig. \ref{ThinDTNG50y100}, galaxies exhibit higher thin disc fractions in the outskirts of clusters in both simulations, compared to those near the centre, across all halo masses. Our rotation-dominated satellite sample thus shows a morphology–environment trend consistent with the classical morphology–density relation. For the total subsamples (gray lines), the thin disc fraction increases from $0.25$ to $0.42$ in TNG50 and from $0.29$ to $0.40$ in TNG100. At fixed cluster-centric distances below $r/R_{200}<1.5$, galaxies in lower-mass haloes also display higher thin disc components. These results indicate that galaxies in less dense environments, such as the outskirts of clusters, tend to preserve higher disc fractions typical of late-type galaxies, whereas in denser environments, closer to cluster centres, galaxies are dominated by lower disc fractions, characteristic of early-type morphologies. This trend can be physically explained by the stronger environmental processes acting in dense cluster regions. Near the cluster centres, galaxies are subjected to mechanisms such as ram-pressure stripping, galaxy harassment, and tidal interactions, which can remove gas reservoirs, suppress star formation, and dynamically heat stellar discs, ultimately driving the transformation toward early-type morphologies \citep{boselli2006environmental}. In contrast, galaxies in the outskirts experience weaker environmental effects, allowing them to retain their gas and sustain star formation, thereby maintaining thin stellar discs and late-type morphologies.

\begin{figure}
\includegraphics[width=\columnwidth]{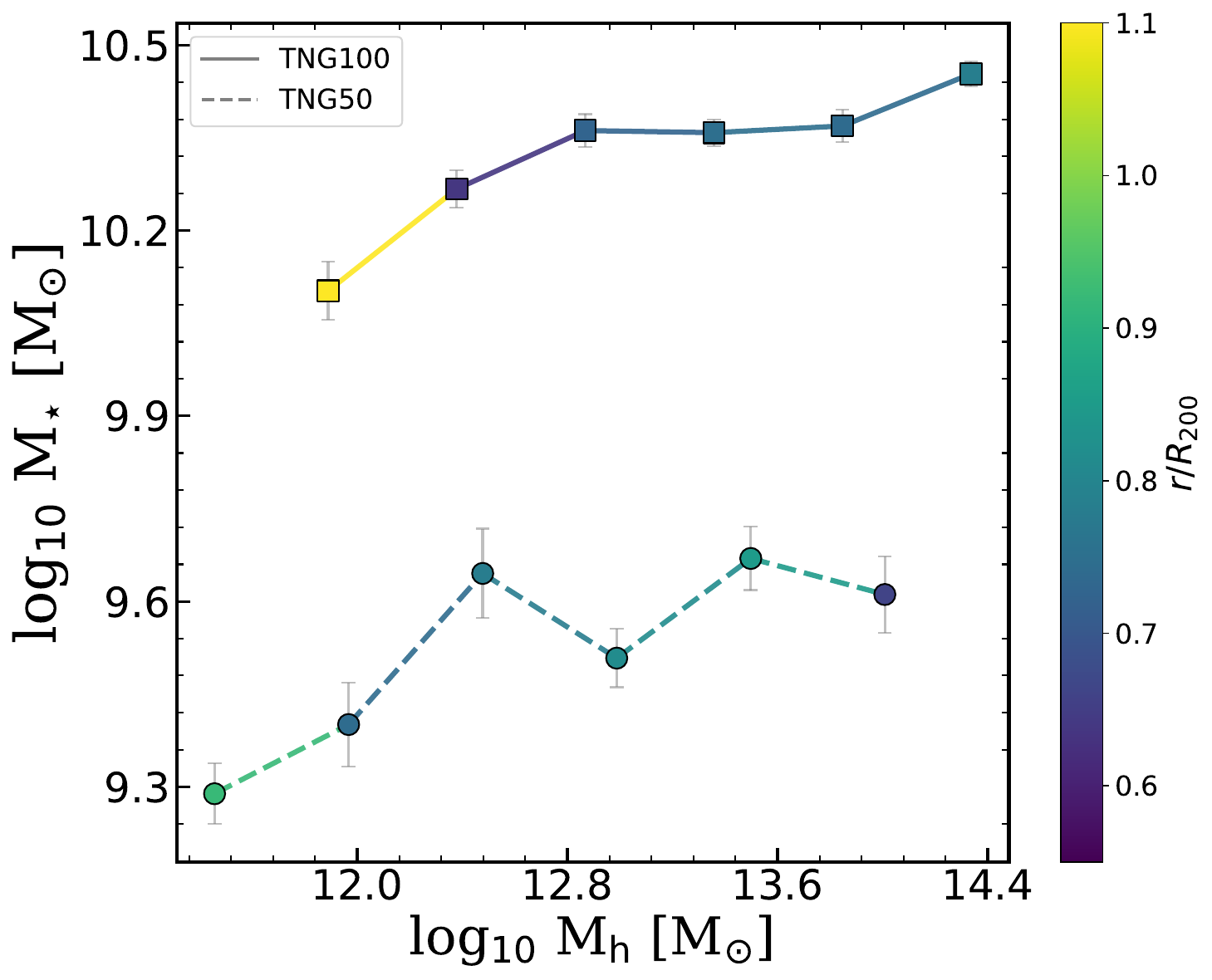} 
\caption{Stellar mass of galaxies as a function of halo mass M$_{\rm h}$ for TNG100 and TNG50, denoted by solid and dotted lines, respectively. The points and lines are coloured according to the median cluster-centric distance. As observed, the environments in which galaxies reside differ between the two simulations; galaxies in TNG100 exhibit higher cluster-centric distances than galaxies in TNG50, but in both simulations we observe that galaxies in more massive haloes also exhibit lower cluster-centric distances. }
\label{logM_vs_M200}
\end{figure}

On the other hand, given that stellar mass is one of the most fundamental parameters of galaxies, in Fig. \ref{logM_vs_M200} we explore how it varies as a function of halo mass M$_{\rm h}$ for satellite galaxies, with colours indicating the median of the cluster-centric distance. As observed in the figure, stellar mass increases with halo mass, and galaxies in more massive haloes tend to lie closer to the cluster centre. It is noticeable also that the stellar mass ranges in TNG50 and TNG100 are different, with TNG50 including lower-mass galaxies (M$_{\star} \geq 10^{9}$ M$_{\odot}$) and TNG100 sampling more massive galaxies (M$_{\star} \geq 10^{10}$ M$_{\odot}$). As a result, both simulations span similar halo mass ranges, including low, intermediate, and high-mass haloes. However, TNG100 contains a larger number of massive haloes, while galaxies in TNG50 are predominantly found in lower-mass haloes. These differences reflect the complementary nature of the two simulations. In the literature, the stellar mass–halo mass relation for satellite galaxies is known to be sensitive to environmental effects. At fixed satellite dynamical (subhalo) mass, satellites residing in more massive hosts and at smaller host-centric radii have larger stellar-to-dynamical mass ratios, consistent with stronger dark-matter stripping \citep{engler2021distinct}. The global relation for the full galaxy population in IllustrisTNG has been characterised by \cite{pillepich2018first}, who showed that total halo mass is a good predictor of stellar mass over a wide range of halo masses, while the properties the stellar and gaseous components evolution of galaxies in TNG50 are studied in \citet{pillepich2019first}. However, the purpose of Fig. \ref{logM_vs_M200} is not to reproduce those global relations, but to show the halo mass and cluster-centric distance distributions of our selected satellite samples.

\subsection{Bar fraction dependence on stellar mass and morphology}

To analyze the bar presence in our final subsamples, we calculated the bar fraction (f$_{\rm bar}$), defined as the ratio between the number of barred disc galaxies and the total number of disc galaxies. For our subsamples of satellite galaxies, we obtain a total bar fraction of $29.22 \% \pm 1.4\%$ (1$\sigma$ bootstrap uncertainties) and $47.46 \% \pm 1.2 \%$ for TNG50 and TNG100, respectively. The lower bar fraction in TNG50 is, at least in part, a consequence of the stellar mass range explored, which extends to lower masses where bars have a lower likelihood of being present. 

The link between bar presence, galaxy morphology, and environment has been widely discussed in the literature. Observational studies have shown that the fraction of barred galaxies is strongly linked to the morphological type and stellar mass of galaxies, being highest among massive late-type systems. For instance, \citet{nair2010fraction} classified late-type galaxies as those with disc-like axial ratios ($b/a > 0.4$) based on their photometric decomposition, while \citet{tawfeek2022morphology} employed \texttt{MORPHOT}, an automated nonparametric tool that applies a neural network to galaxy images to improve the classification between elliptical and S0 galaxies, and defined late-type galaxies as those with morphological type $M_{\rm type} > 0$, corresponding to disc-dominated systems. \citet{nair2010fraction} reported that the bar fraction is sensitive to both stellar mass and central concentration, indicating that bars are more frequent in disc-dominated galaxies with active star formation, whereas they are scarce in more bulge-dominated or quenched systems. On the other hand, \citet{tawfeek2022morphology} focused on how the connection between bar presence and morphology is modulated by environmental effects, finding that the bar fraction decreases toward the centres of massive clusters, but this trend largely disappears once the morphological type is controlled for. This suggests that the likelihood of a galaxy hosting a bar is primarily driven by its morphological transformation, a result that motivates our exploration of the morphology–density relation in our samples. Observational studies have shown that the relationship between bar fraction and galaxy morphology is complex and depends on both stellar mass and bar strength. For massive galaxies (M$_{\star} \geq 10^{11}$ M$_{\odot}$), the bar fraction tends to be lower in early-type discs than in late-type discs \citep{cameron2010bars}, although this trend reverses at intermediate masses. Furthermore, when considering only strong bars, these are more common in early-type spirals, whereas weak bars are more frequent in late-type spirals \citep{lee2019bar}. Given that early-type morphologies are more common in dense environments, and that the presence of thin discs decreases towards smaller cluster-centric distances, we would expect a decline of the bar fraction for decreasing cluster-centric distance in our samples.

\begin{figure}
\includegraphics[width=\columnwidth]{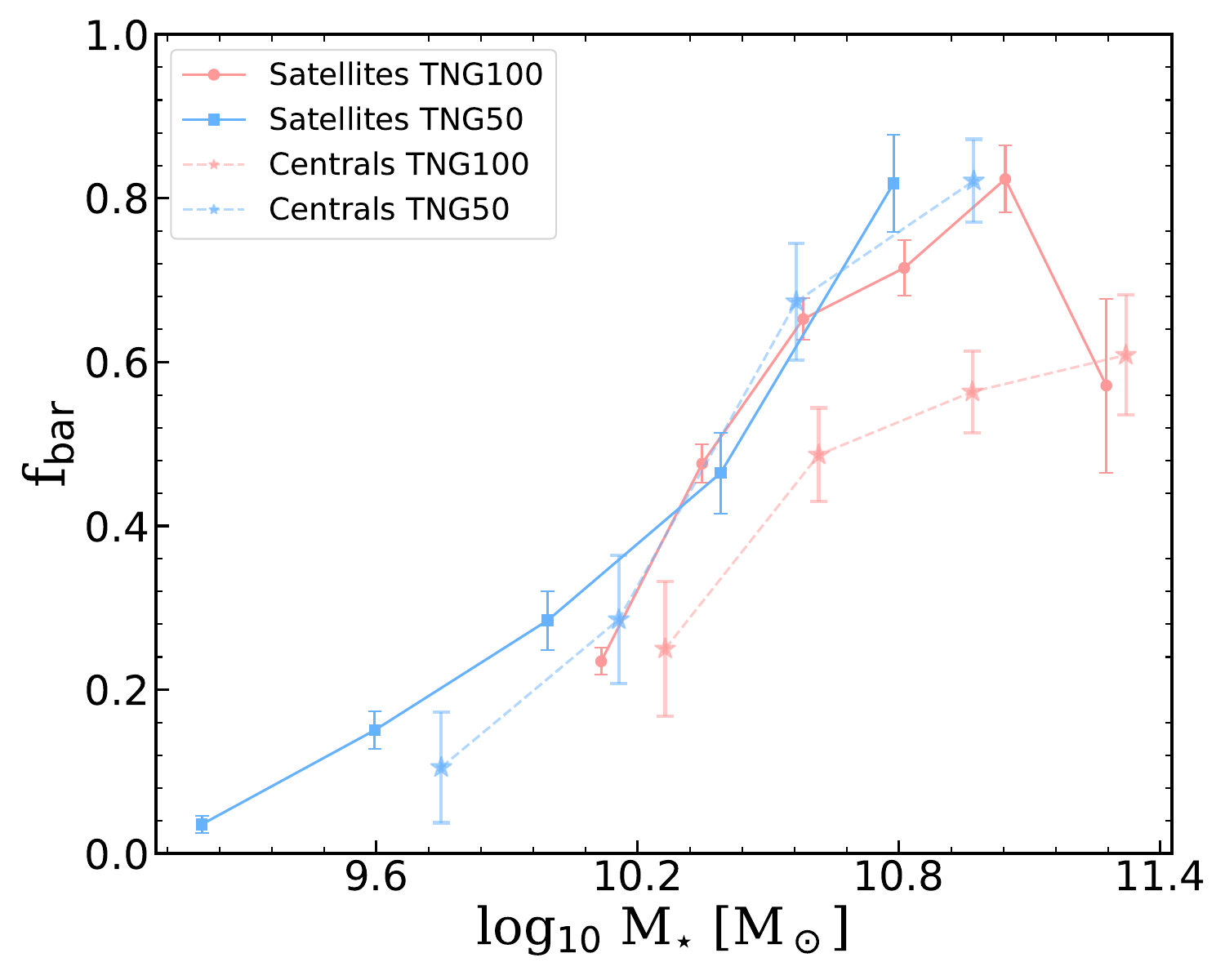} 
\caption{Bar fraction as a function of stellar mass for both simulations, TNG50 (blue lines) and TNG100 (pink lines), at $z=0$, with central and satellite galaxies denoted by dotted and solid lines, respectively. We observe that bar presence increases with stellar mass in both simulations, and that there is no difference in bar fraction between the simulations in the stellar mass range $10^{10}$–$10^{11.2} \rm M_{\odot}$ for satellite galaxies. Central galaxies in TNG100 clearly exhibit lower bar fractions.}
\label{fbar_logM_50_100}
\end{figure}

In Fig. \ref{fbar_logM_50_100} we observe the behaviour of the bar fraction f$_{\rm bar}$ as a function of the stellar mass for central (dotted lines) and satellite (solid lines) galaxies. As shown, for all samples, the bar fraction increases with stellar mass in both simulations. For galaxies with stellar masses above $\sim 10^{10.2}$ M${\odot}$, the bar fraction is similar between satellites in TNG100 and TNG50. For satellites in TNG50, f$_{\rm bar}$ rises from $\sim 5 \%$ for the least massive galaxies to $\sim 75 \%$ for the most massive ones, while for satellites in TNG100 this fraction increases from $\sim 15 \%$ to $\sim 60 \%$. In contrast, the bar fraction for central galaxies in TNG100 is noticeably lower, although it also increases with stellar mass. The trend of increasing bar fraction with stellar mass has also been reported in several observational (e.g., \citealt{sheth2008evolution, cervantes2017stellar}) and theoretical (e.g., \citealt{peschken2019tidally, rosas2020buildup}) studies. However, \citet{erwin2018dependence} showed that the bar fraction declines above  $10^{10}$ M${_\odot}$ when correcting for angular resolution, which differs from our results, where the decline occurs above  $10^{11}$ M${_\odot}$. This difference can be explained by the results of \citet{zhao2020barred}, who found that TNG100 overproduces short bars (radius  $ \sim 1.4$--$3$ kpc) in massive galaxies, and that these short bars are responsible for the decline at high masses. Similarly, \citet{lu2025illustristng} found that in TNG50 the bar fraction flattens at $10^{10}$ M${_\odot}$ and decreases at higher masses, and also noted that IllustrisTNG simulations produce an excess of short-bar galaxies. Given that we include the full bar population regardless of bar size, our decline in bar fraction occurs at higher stellar masses.

\begin{figure*}
\captionsetup[subfigure]{labelformat=empty}
\centering
\begin{subfigure}{0.5\linewidth}
    \includegraphics[width=\linewidth]{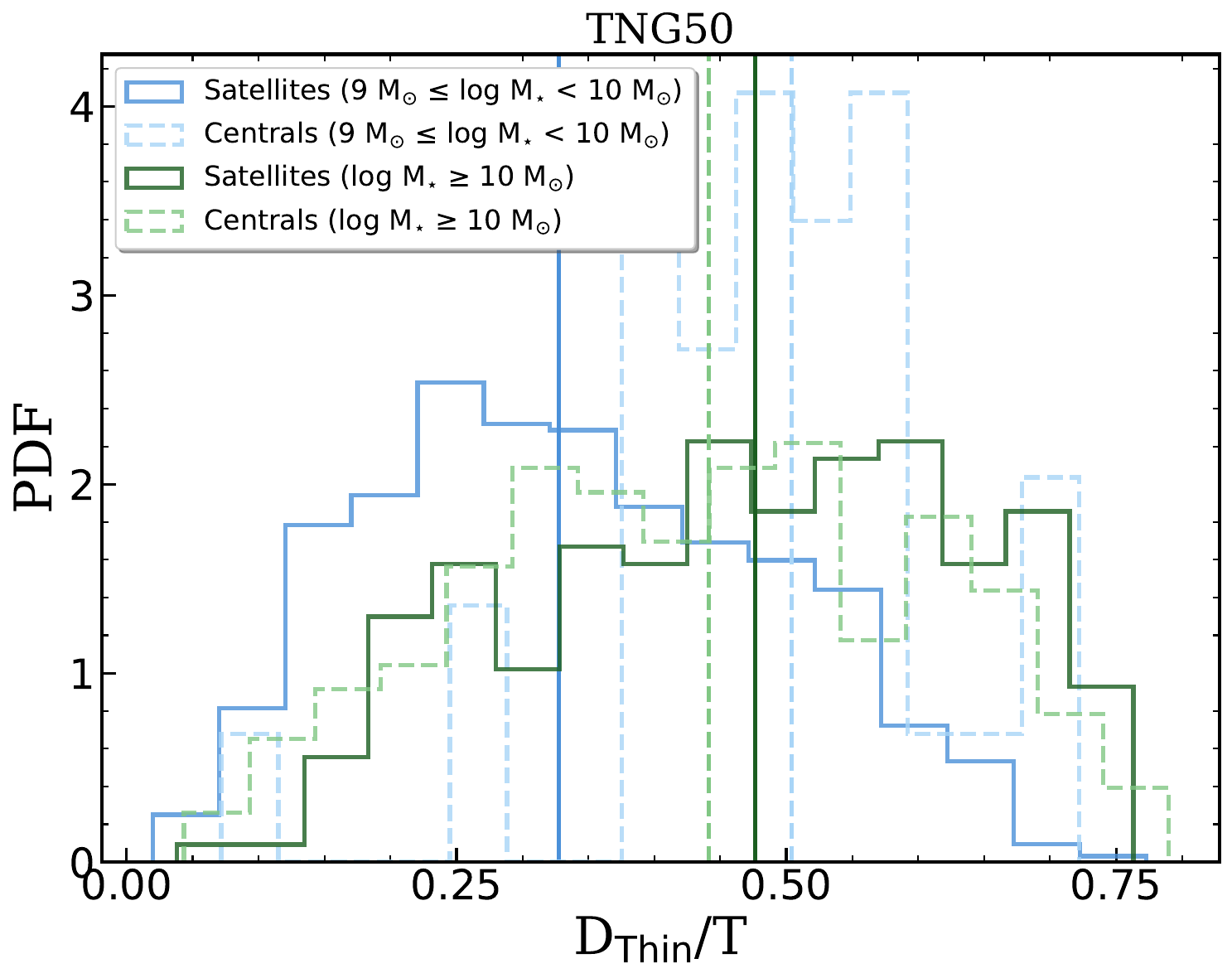}
    \subcaption{}
\end{subfigure}
\hspace{-0.01\linewidth}
\begin{subfigure}{0.5\linewidth}
    \includegraphics[width=\linewidth]{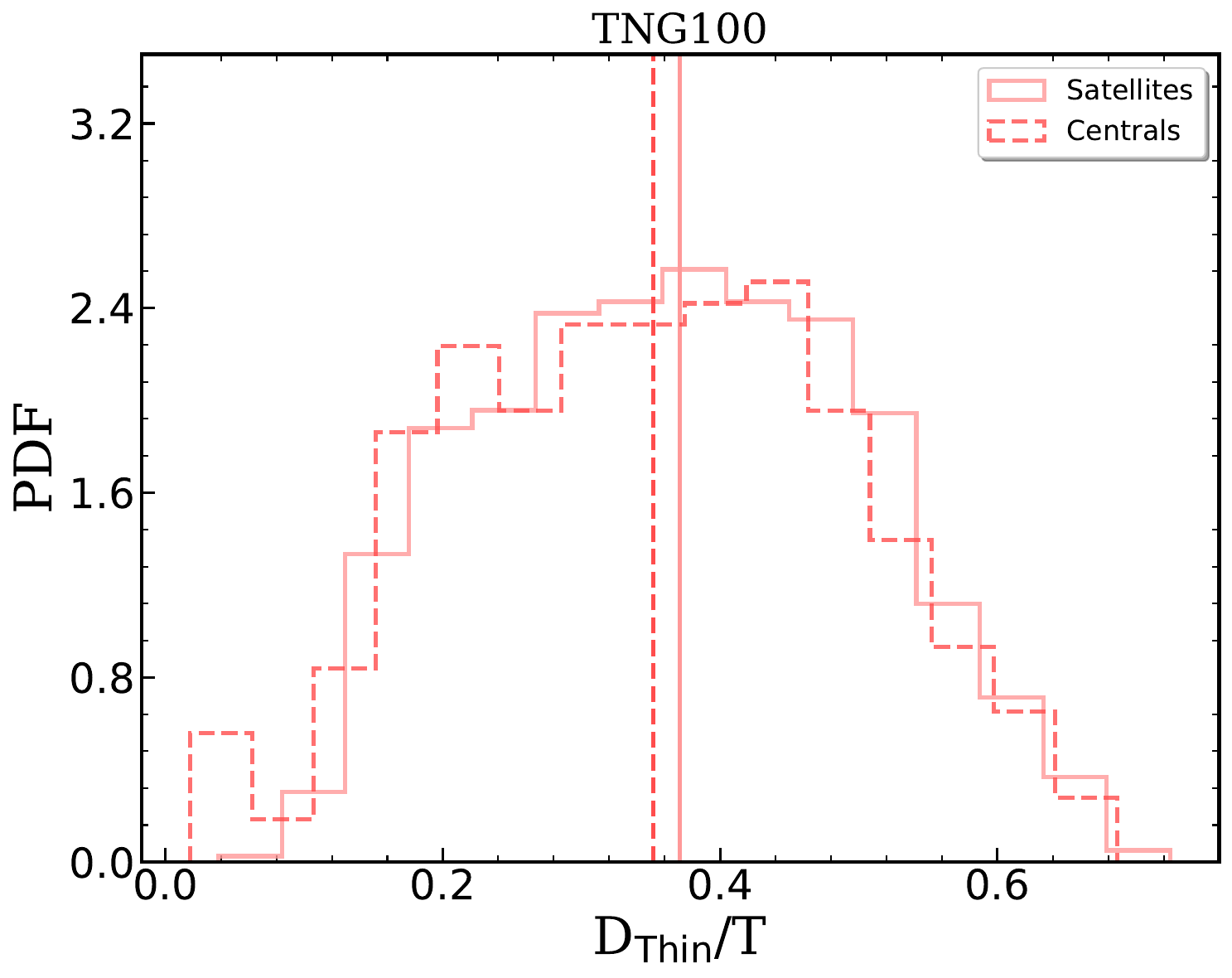}
  \subcaption{}
\end{subfigure}
\caption{Probability density function of the thin disc component for TNG50 (left panel) and TNG100 (right panel). In these panels, we observe that central galaxies in TNG100 exhibit lower thin disc components than satellites, suggesting that these galaxies are less disky. The opposite trend is observed for TNG50, where satellites show lower thin disc components for galaxies with stellar masses lower than $10^{10}$ M$_{\odot}$.} 
\label{ThinDTNGs}
\end{figure*}

A possible explanation for the lower bar fraction observed in central galaxies compared to satellites in TNG100, as well as for the differences between the two simulations, may lie in the different distributions of thin disc fractions in these populations. To explore this, in Fig. \ref{ThinDTNGs} we present the thin disc distribution for both simulations, separating central and satellite galaxies. As shown, in TNG100 (right panel), central galaxies exhibit slightly lower thin disc components, with a median value of $0.35$, compared to satellites with $0.37$. However, this difference is small and does not appear sufficient to fully account for the lower bar fraction observed in central galaxies. In TNG50 (left panel), we find the opposite trend when considering the stellar mass range $10^9-10^{10}$ M$_{\odot}$, with central galaxies exhibiting higher thin disc parameters than satellites (median values of $0.36$ and $0.50$ for satellites and centrals, respectively), unlike in TNG100. When restricting the analysis to the same stellar mass range as in TNG100, we recover the trend that satellite galaxies exhibit higher thin disc components.

\begin{figure*}
\captionsetup[subfigure]{labelformat=empty}
\centering
    \begin{subfigure}{0.5\linewidth}
        \includegraphics[width=\linewidth]{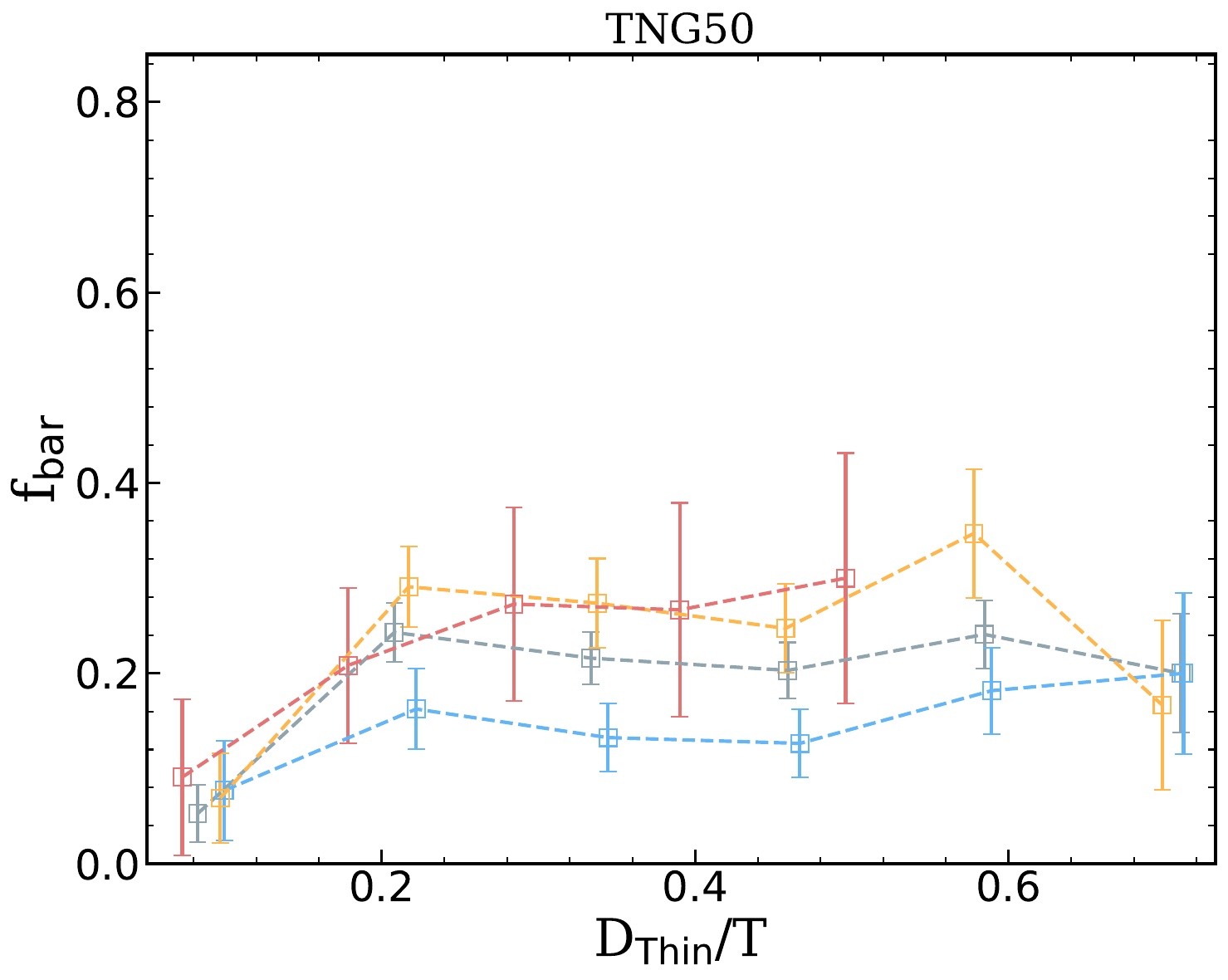}
        \subcaption{ }
        \label{fig:sfig1}
    \end{subfigure}
    \hspace{-0.01\linewidth}
    \begin{subfigure}{0.5\linewidth}
        \includegraphics[width=\linewidth]{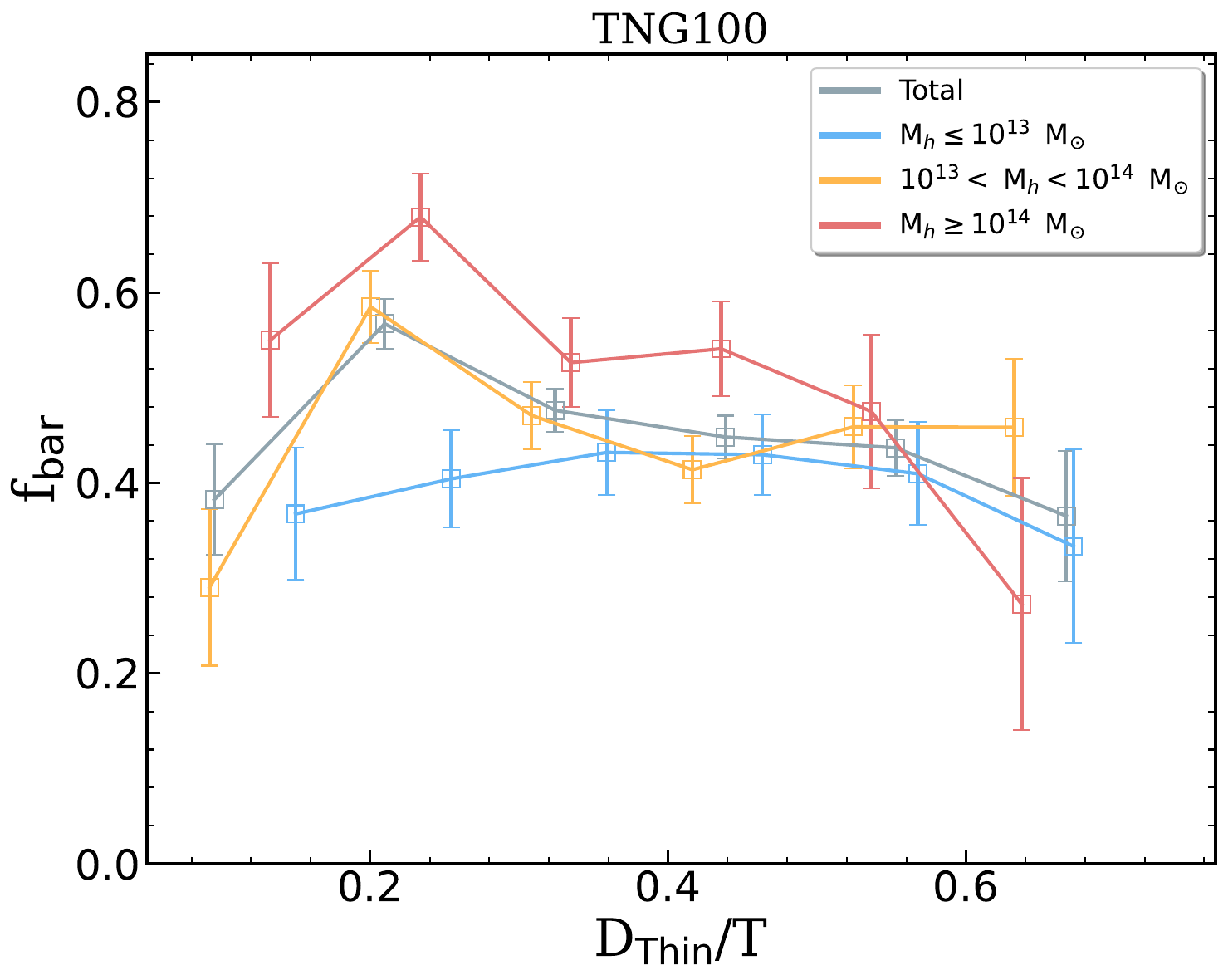}
        \subcaption{ }
        \label{fig:sfig2}
    \end{subfigure}

\caption{Bar fraction as a function of the thin disc component for TNG50 (left panel) and TNG100 (right panel). In both simulations, the bar fraction increases for galaxies with a more prominent thin disc. At a fixed thin disc component, the bar fraction is higher in the more massive haloes in TNG100, while in TNG50 this trend is evident for the low- and intermediate-mass haloes.} 
\label{fbar_Dthin}
\end{figure*}

In Fig. \ref{fbar_Dthin} we observe the bar fraction as a function of the thin disc mass fraction D$_{\rm Thin}$/T, for the satellite galaxies in both simulations and for the different halo masses. From the figure, we find that the bar fraction increases only at small thin disc fractions, while at intermediate and larger values the trend flattens or even decreases. For the total samples, the bar fraction rises from $\sim 5 \%$ to $\sim 24 \%$ in TNG50 for galaxies with a thin disc component lower than $0.20$; beyond this value, the bar fraction remains approximately constant at intermediate and larger thin disc fractions. This behaviour is consistent with \cite{DuRevisiting2026}, who studied the excess of bar-like structures in TNG50 early-type galaxies (ETGs) and showed that these features are not determined solely by the thin disc component. Instead, their abundance is also linked to the dynamical state of the galaxies, particularly their low spin and gas-poor nature, which remove the stabilizing influence of a cold, high-angular-momentum gas disc. In this picture, many of these structures are long-lived, secularly evolved remnants of earlier bars that can persist even in dispersion-dominated systems, naturally leading to a weak or flat dependence of $f_{\rm bar}$ on the thin disc component. On the other hand, for TNG100 the bar fraction for the total sample increases from $\sim 36 \%$ to $\sim 56 \%$ for galaxies with thin disc component below $0.20$. Beyond this threshold, the bar fraction shows a clear decline for the most massive haloes, while for intermediate-mass haloes it decreases at intermediate thin disc fractions and stabilises at larger values, and for low-mass haloes it decreases at the largest thin disc fractions. Regarding the halo masses at fixed thin disc mass fraction, more massive haloes tend to present higher bar fraction in both simulations.

\subsection{Bar properties in different environments.}
\begin{figure*}
\captionsetup[subfigure]{labelformat=empty}
\centering
    \begin{subfigure}{0.47\linewidth}
        \includegraphics[width=\linewidth]{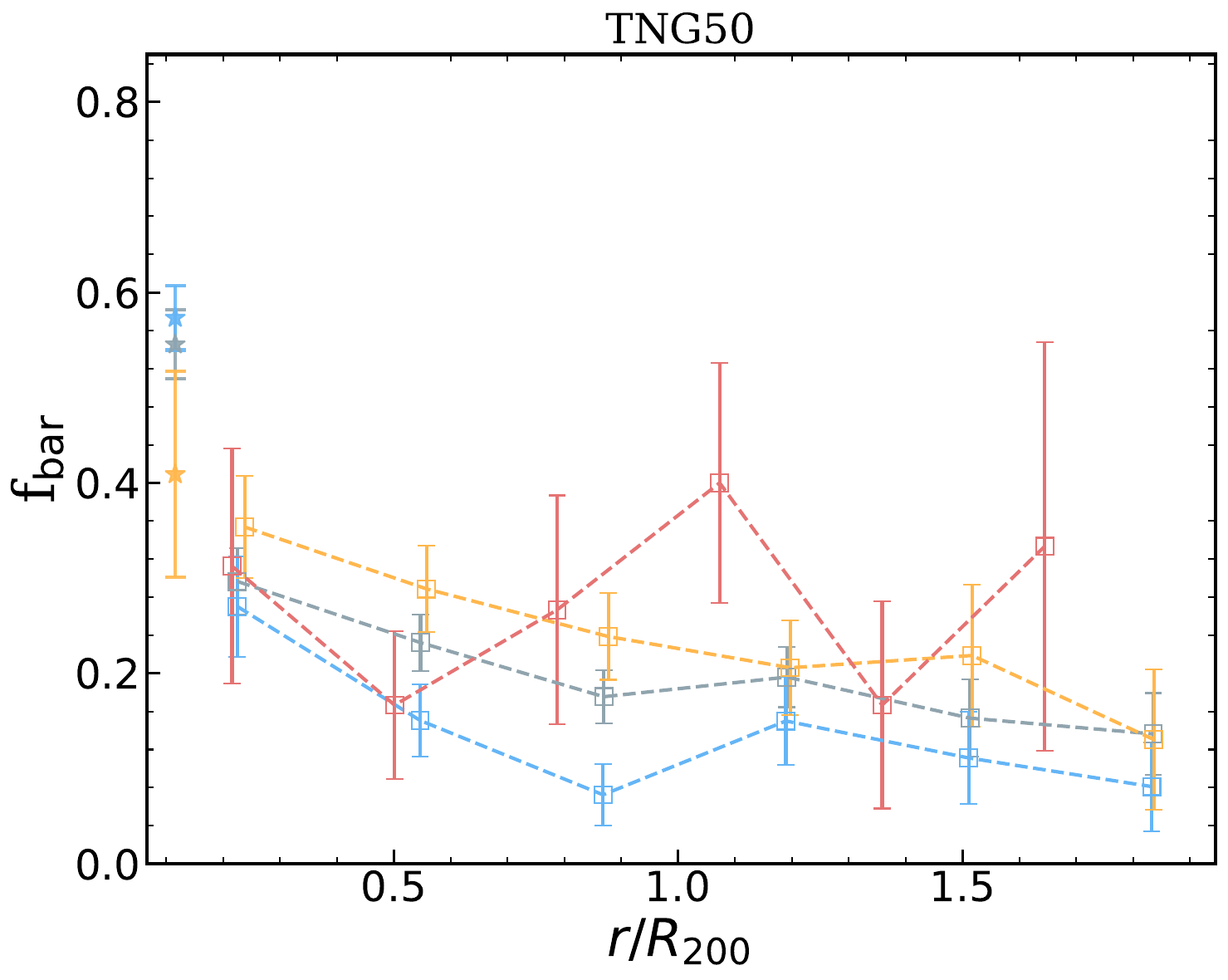}
        \subcaption{ }
        \label{fig:sfig1}
    \end{subfigure}
    \hspace{-0.01\linewidth} 
    \begin{subfigure}{0.47\linewidth}
        \includegraphics[width=\linewidth]{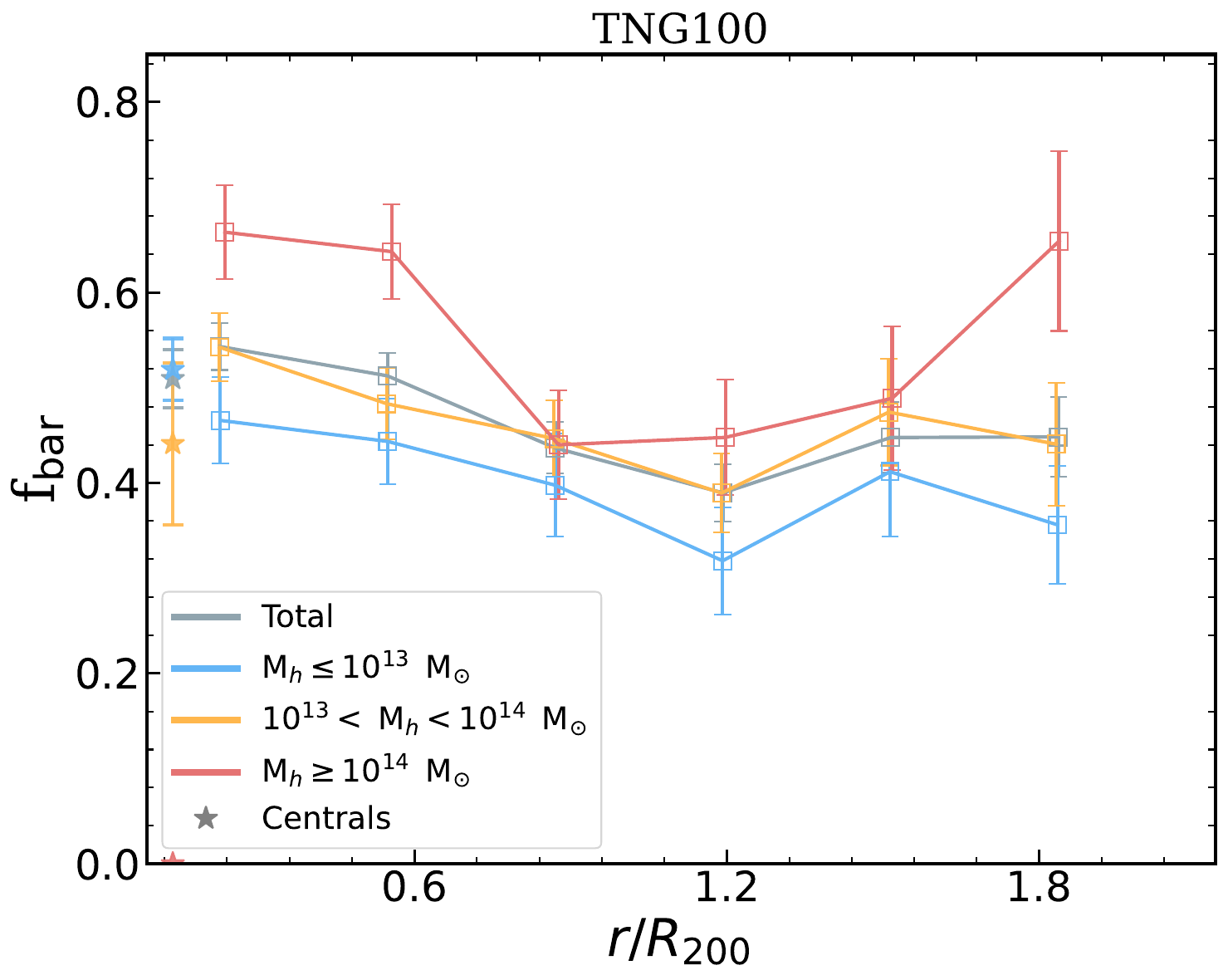} 
        \subcaption{ }
        \label{fig:sfig2}
    \end{subfigure}

    \vspace{-0.2cm} 

    \begin{subfigure}{0.47\linewidth}
        \includegraphics[width=\linewidth]{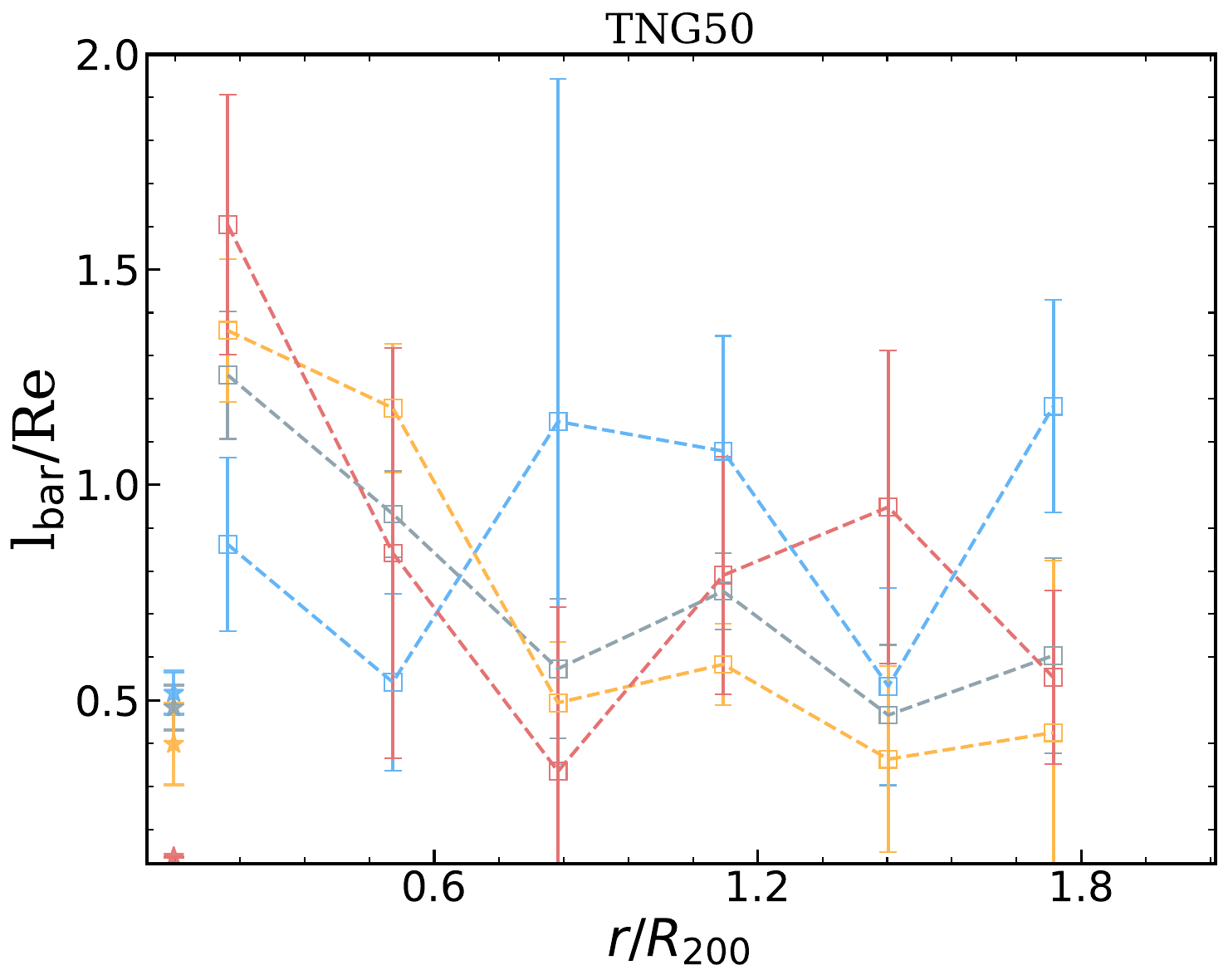}
        \subcaption{ }
        \label{fig:sfig3}
    \end{subfigure}
    \hspace{-0.01\linewidth}
    \begin{subfigure}{0.47\linewidth}
        \includegraphics[width=\linewidth]{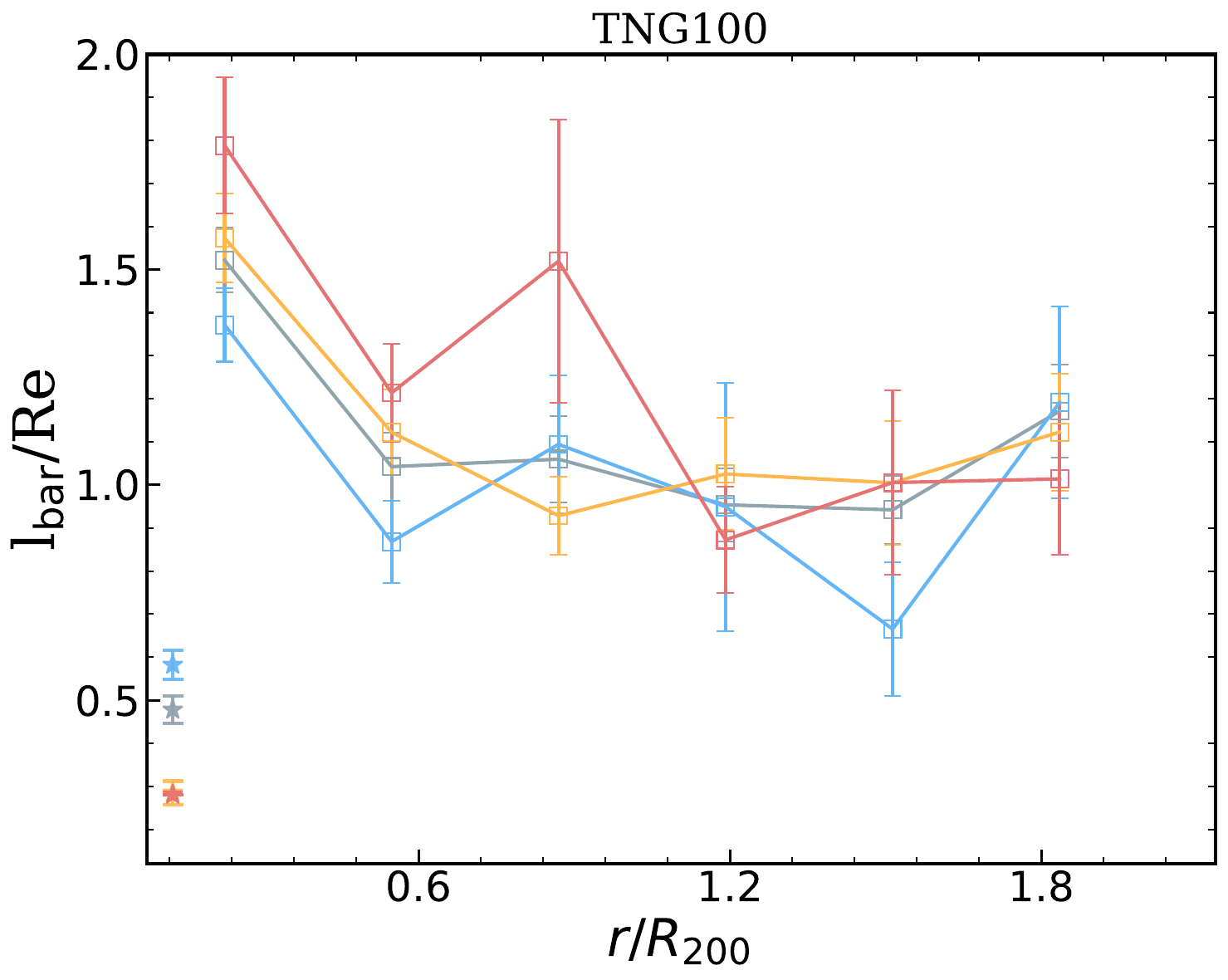}
        \subcaption{ }
        \label{fig:sfig4}
    \end{subfigure}

    \vspace{-0.2cm} 

    \begin{subfigure}{0.47\linewidth}
        \includegraphics[width=\linewidth]{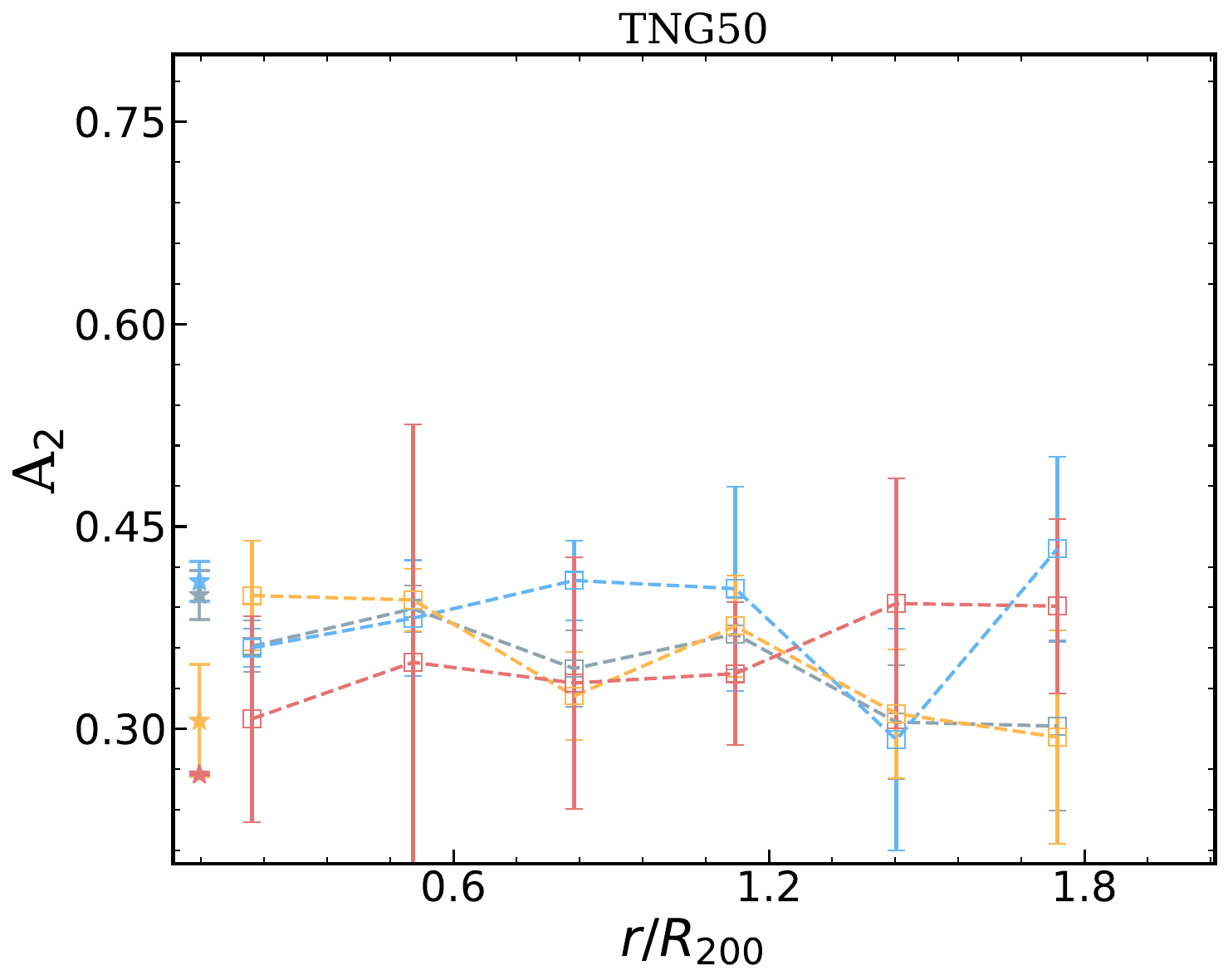}
        \subcaption{ }
        \label{fig:sfig3}
    \end{subfigure}
    \hspace{-0.01\linewidth}
    \begin{subfigure}{0.47\linewidth}
        \includegraphics[width=\linewidth]{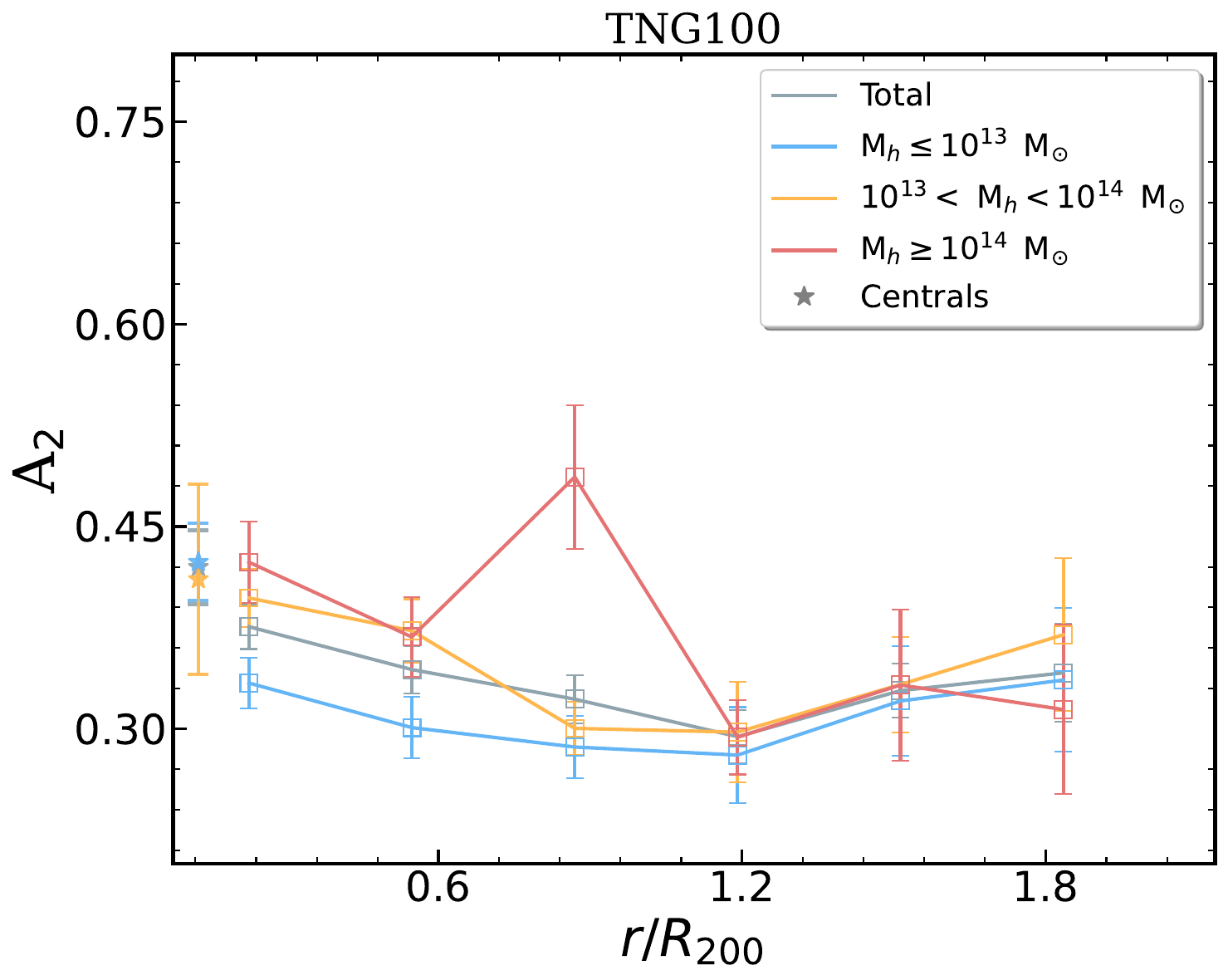}
        \subcaption{ }
        \label{fig:sfig4}
    \end{subfigure}

\caption{Bar properties as a function of cluster-centric distance for different halo masses. The upper panels show the bar fraction, which decreases with increasing distance. The second row shows the normalized bar radius (normalized to the effective radius), while the lower panels show the bar strength A$_2$. As we observe, the normalized sizes are larger closer to the cluster centre. Bars also tend to be stronger in these regions. The star symbols represent the median values for centrals.} 
\label{bar_prop}
\end{figure*}

In the previous section, we examined how the bar fraction depends on internal parameters such as stellar mass and the thin disc component. Here, we explore how bar properties vary with the environment, accounted for by the membership to groups/clusters of different total masses. In Fig. \ref{bar_prop} we show the bar fraction (upper panels), normalized bar radius l$_{\rm bar}/R_{e}$ (middle panels), and bar strength $A_2$ (lower panels) as a function of cluster-centric distance, split by halo mass, for TNG50 (left) and TNG100 (right). In the upper panels we observe a slight decline of the bar fraction from the centre of the clusters towards the outskirts in both simulations: at the centres the bar fraction reaches $\sim 33 \%$ in TNG50 and $\sim 54 \%$ in TNG100, while in the outermost regions it falls to $\sim 22 \%$ and $\sim 44\%$, respectively. This suggests that bars are found preferably in galaxies with lower cluster-centric distances.

The slight decrease in bar fraction towards the cluster outskirts that we find is consistent with some observational studies reporting higher bar fractions in cluster centres. For example, \cite{thompson1981bar} showed that strong tidal interactions in the Coma cluster can trigger disc instabilities leading to bar formation in otherwise stable galaxies, while \cite{andersen1996distribution} found a similar enhancement of barred spirals in the central regions of Virgo. However, other works argue that the bar fraction does not vary significantly with cluster-centric distance. Studies such as \cite{mendez2010lack}, \cite{Mendez2023JWST}, and \cite{lansbury2014barred} concluded that repeated interactions heat stellar discs, preventing bar formation, and that long-lived strong bars should remain largely unaffected by the environment. More recently, \citet{tawfeek2022morphology} reported the opposite trend, finding that the bar fraction decreases toward the centres of clusters in the OmegaWINGS sample. They attributed this behaviour to the morphological transformation of galaxies, showing that once morphological type is controlled for, the dependence of the bar fraction on cluster-centric distance vanishes. Given that our simulated galaxy sample reproduces the observed morphology–density relation, one would expect a similar trend to emerge in our analysis. However, this is not the case: despite the consistency of our simulations with the environmental segregation of morphologies, we still find a slight decrease in the bar fraction toward the cluster outskirts. Theoretical studies also reach mixed conclusions: while some zoom-in simulations suggest that bars form primarily due to intrinsic disc instabilities, largely independent of interactions (\citealt{zana2018external}), other N-body works show that tidal encounters can induce earlier and stronger bar formation in dense environments (\citealt{lokas2016tidally, pettitt2018bars}). 

In addition to the bar fraction behaviour shown in these panels, we also note that at a fixed cluster-centric distance, the bar fraction is higher in more massive haloes. This trend is particularly evident in TNG100: across the entire $r/R_{200}$ range, haloes with M$_{\rm h} \geq 10^{14} {\rm M}_{\odot}$ exhibit higher bar fractions than intermediate-mass haloes, which in turn show higher fractions than low-mass haloes. In TNG50, the trend is more apparent between intermediate- and low-mass haloes; for the single high-mass halo available in this simulation, the behaviour appears erratic, likely due to the limited number of galaxies and poor statistics. To better understand the origin of these differences, it is therefore important to examine the properties of galaxies in the different halo masses, such as their total stellar masses and stellar assembly times.

The middle and lower panels of Fig. \ref{bar_prop} show the normalized bar radius (relative to the effective radius) and the bar strength of barred galaxies. As observed, the normalized size decreases with cluster-centric distance in both simulations. This behaviour can be explained by the fact that galaxies located near the cluster centre tend to be smaller than those in the outskirts. As a consequence, the normalized bar radius decreases with distance. A similar trend in the distribution of galaxy sizes within clusters has been reported by \citet{kuchner2017effects}, who found that cluster galaxies closer to the centre exhibit a smaller effective radius compared to those at larger cluster-centric distances. Consistently, \citet{aguerri2023properties} find that galaxies in the virialized and infall regions of the Virgo cluster exhibit smaller sizes compared to field galaxies. This is also consistent with \citet{Cuomo2025prop}, who find that bars in high-density environments, such as galaxies in the Virgo cluster, are shorter than those in filaments, and even shorter compared to those in the field. In TNG100, we also observe that within one virial radius, at fixed cluster-centric distance, bars in more massive haloes exhibit larger normalized sizes. The corresponding behaviour of the bar radius as a function of cluster-centric distance is presented in Appendix~\ref{AppendixA}. As observed, bar sizes remain constant across all cluster-centric distance ranges in both simulations.

Finally, regarding bar strength (A$_2$), for the total subsamples we find a slight decrease in A$_2$ as a function of cluster-centric distance. This trend could be related to the mild decline in the bar fraction, suggesting that bars tend to be somewhat weaker toward the cluster outskirts. 

\subsection{Mass assembly behaviour.}

\begin{figure*}
\captionsetup[subfigure]{labelformat=empty}
\centering
\begin{subfigure}{0.5\linewidth}
    \includegraphics[width=\linewidth]{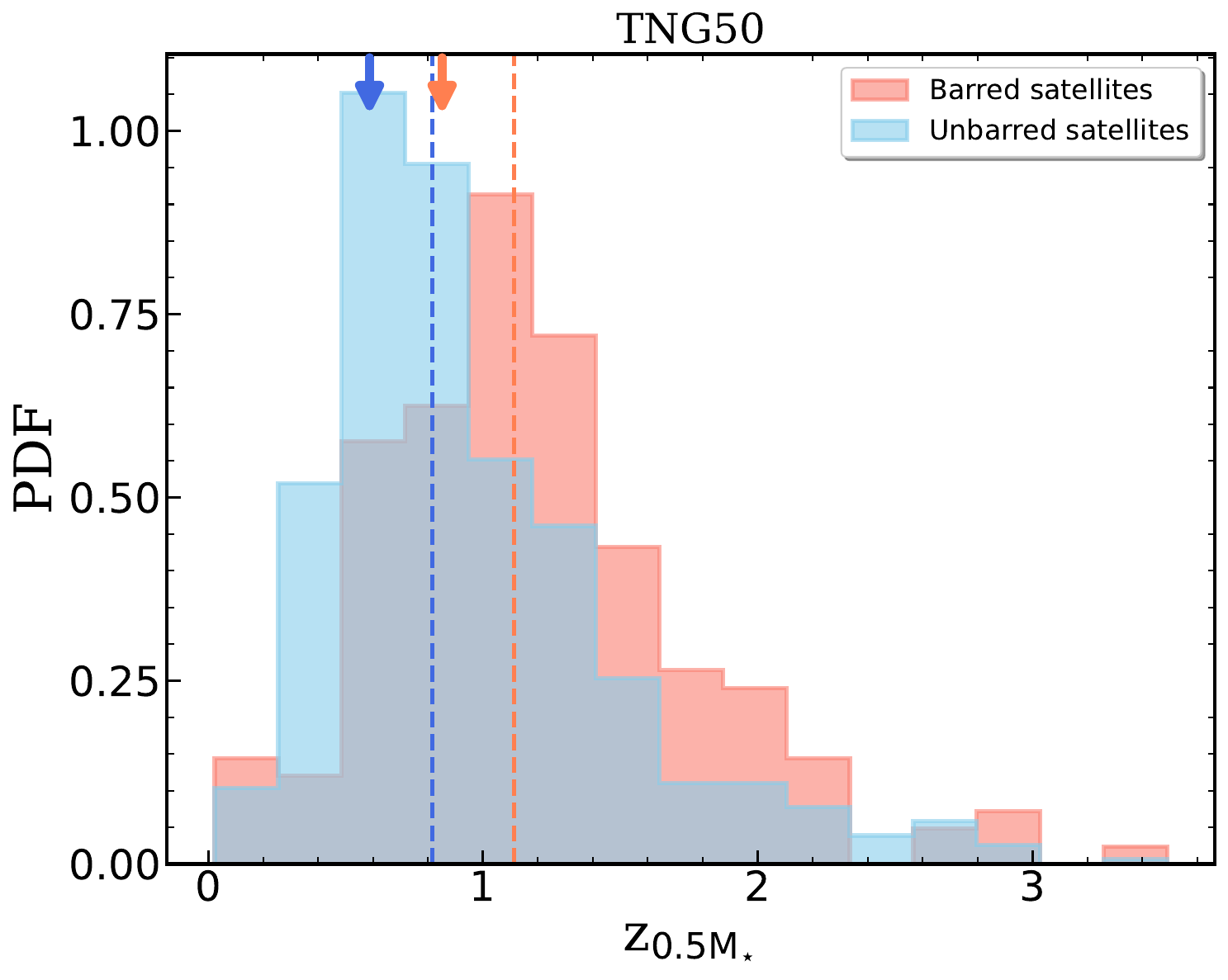}
    \subcaption{}
\end{subfigure}
\hspace{-0.01\linewidth}
\begin{subfigure}{0.5\linewidth}
    \includegraphics[width=\linewidth]{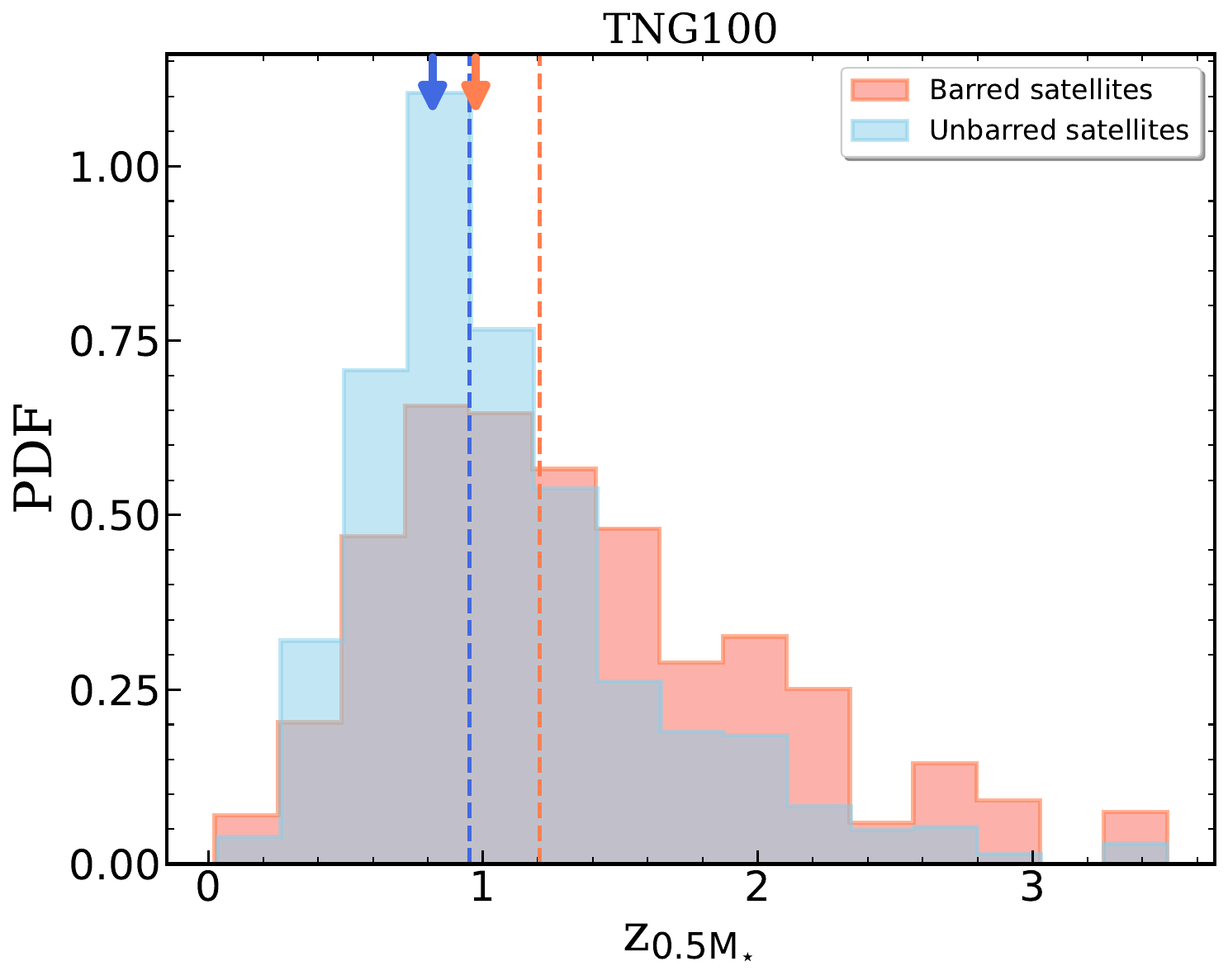}
  \subcaption{}
\end{subfigure}
\caption{Probability density function of the redshift at which galaxies assemble half of their stellar mass for satellite galaxies within 2 virial radii, for TNG50 (left) and TNG100 (right). As observed, barred galaxies exhibit higher assembly redshifts compared to unbarred ones, suggesting that they assembled earlier and could have more time to develop their bars. The vertical dashed lines mark the median of each distribution, while the arrows denote the median values for central galaxies.}
\label{TNG100_50_ASSEMBLY}
\end{figure*}

In this section, we discuss the relationship between the mass assembly of galaxies and bar presence for different environments and halo masses. In Fig. \ref{TNG100_50_ASSEMBLY}, we show the distribution of stellar mass assembly redshift for barred (salmon histograms) and unbarred galaxies (blue histograms) in TNG50 (left) and TNG100 (right). The dotted lines indicate the median values of each distribution. The arrows indicate the median values for central galaxies. In both simulations, barred galaxies exhibit higher median z$_{0.5 \rm M_{\star}}$ values than their unbarred counterparts. For TNG100, the median value of z$_{0.5 \rm M_{\star}}$  for satellite barred and unbarred galaxies is $1.20$ and $0.95$, respectively while for TNG50 is $1.11$ and $0.81$, although the scatter in the distribution is high. These results indicate that barred galaxies tend to assemble their stellar mass earlier than unbarred galaxies, consistent with a scenario in which earlier assembly leads to dynamically more evolved and stable stellar discs at earlier times, making them more susceptible to bar instabilities. This is consistent with \cite{Khoperskov2024why}, who found using TNG50 that the strength of bars at  $z = 0$ correlates more crucially with the time of onset of the stellar disc than with the total mass build-up alone: discs of strongly barred galaxies form early, compared to weakly and non-barred galaxies.

This result is consistent with the broader picture of galaxy evolution commonly referred to as cosmic downsizing, in which more massive galaxies tend to assemble their stellar mass earlier than less massive systems. Observationally, \cite{sheth2008evolution} found that more massive galaxies tend to host bars that formed at higher redshifts, suggesting a connection between the early assembly of massive galaxies and the early formation of bars. Similarly, using the TNG50 simulation, \cite{anderson2024interplay} showed that bars that form earlier have more time to evolve, potentially leading to the development of boxy/peanut structures and affecting the relation between bar radius and galaxy size. In this context, our results suggest that the higher bar fractions associated with earlier stellar mass assembly may reflect a broader connection between galaxy mass, assembly history, and the subsequent evolution of stellar bars.

\begin{figure*}
\centering
\captionsetup[subfigure]{labelformat=empty}
\begin{subfigure}{0.5\linewidth}
    \includegraphics[width=\linewidth]{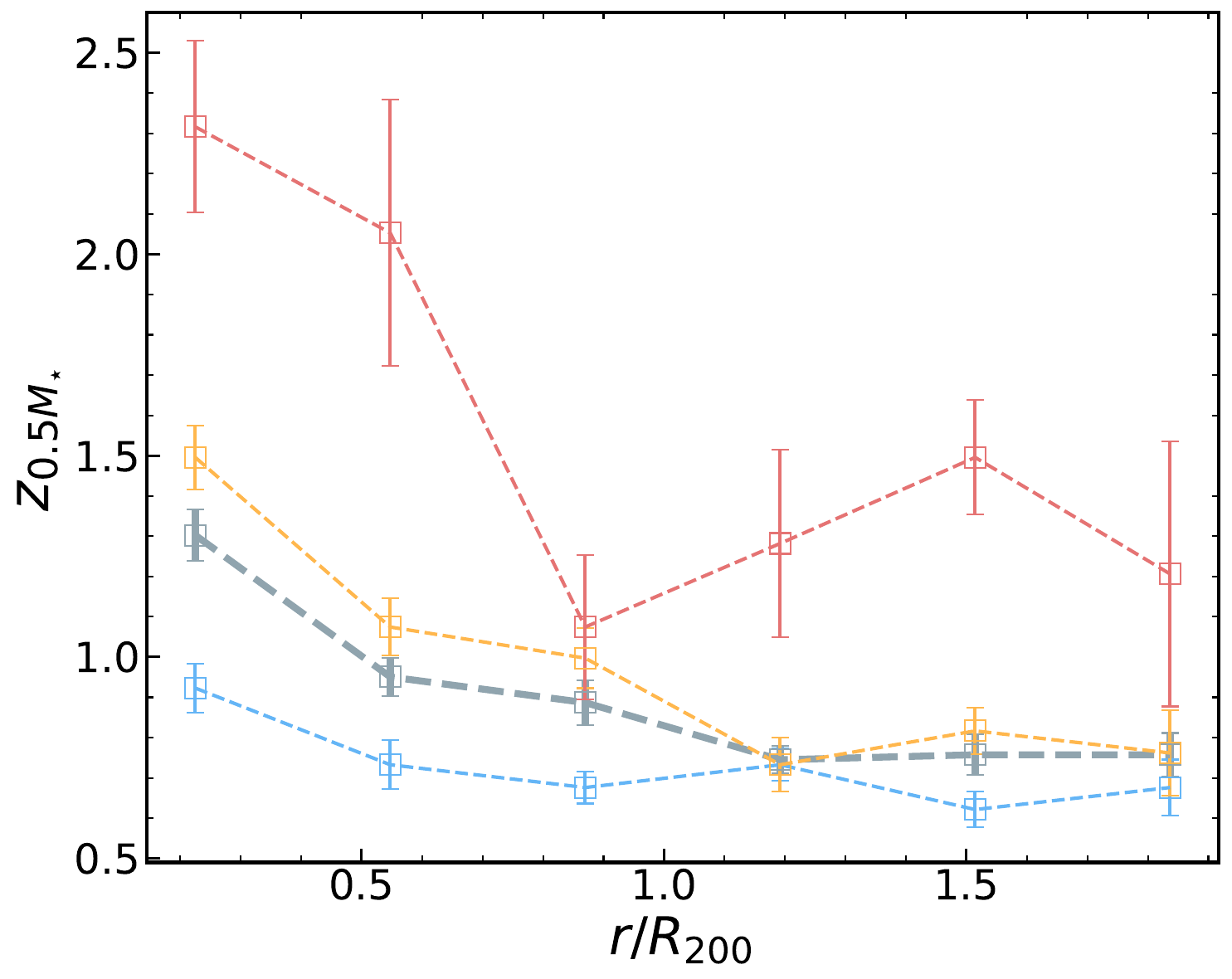}
    \label{05Ms}
\end{subfigure}
    \hspace{-0.01\linewidth}
\begin{subfigure}{0.5\linewidth}
    \includegraphics[width=\linewidth,]{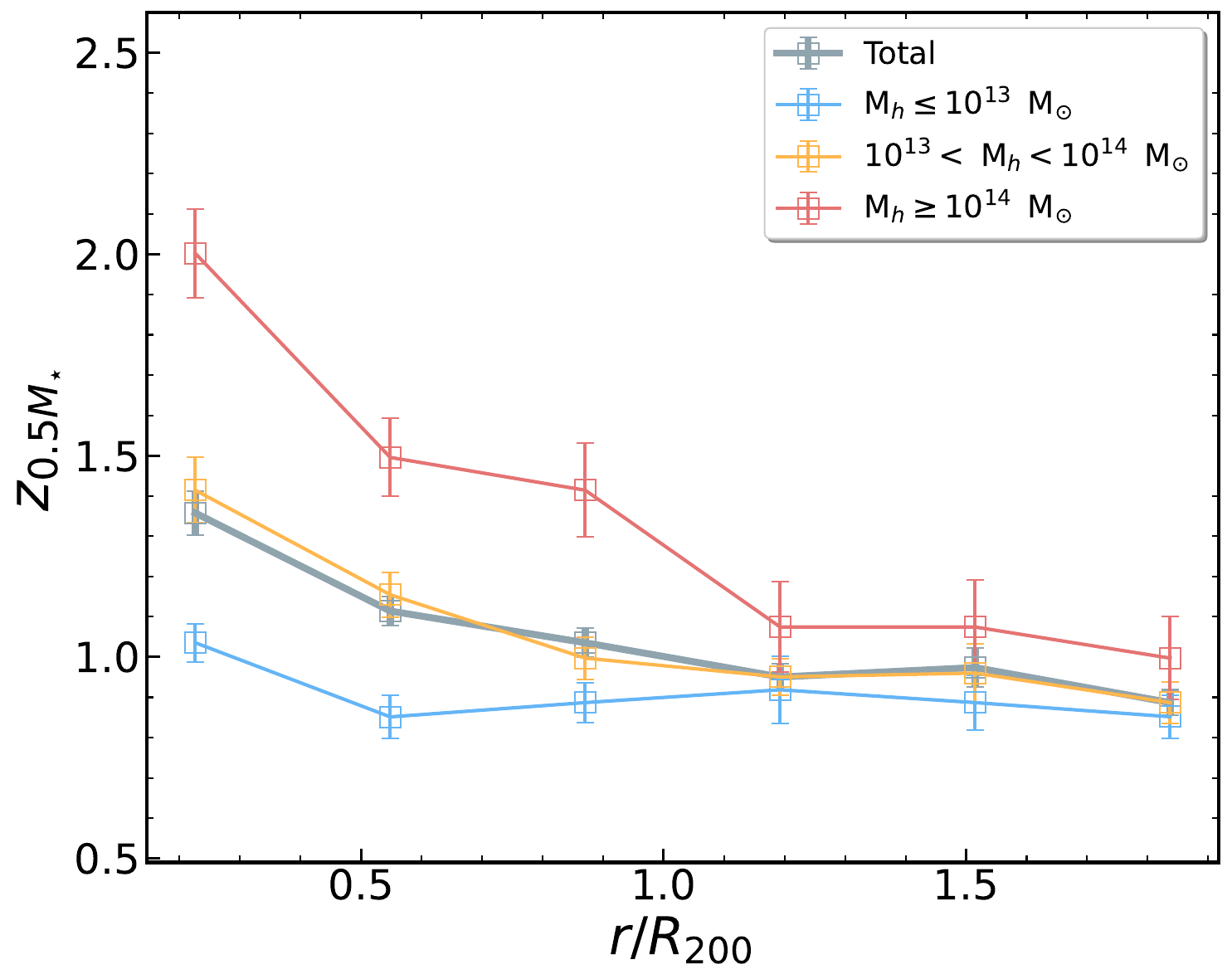}
  \label{05Mdm}
\end{subfigure}
\caption{Mass assembly redshifts as a function of the cluster-centric distance for both simulations, TNG50 (left) and TNG100 (right).} 
\label{z05logMs_vs_rR200}
\end{figure*}

To analyse the assembly time across environments, we plot the stellar mass assembly redshift as a function of cluster-centric distance for haloes of different masses in Fig. \ref{z05logMs_vs_rR200}. We find that the total samples show a clear radial gradient, while individual halo-mass bins are noisier. This behaviour is consistent across the explored halo mass bins. Rather than implying a direct causal dependence on cluster-centric distance, this trend likely reflects the earlier infall and longer residence times of galaxies in the inner regions of clusters. In this context, the observed gradient in assembly history provides a useful framework for interpreting the bar fraction, as the environmental processes that shape galaxy evolution in dense regions may also influence the formation and survival of bars.

\begin{figure*} 
\captionsetup[subfigure]{labelformat=empty}
\begin{subfigure}{0.5\linewidth}
    \centering
    \includegraphics[width=\linewidth]{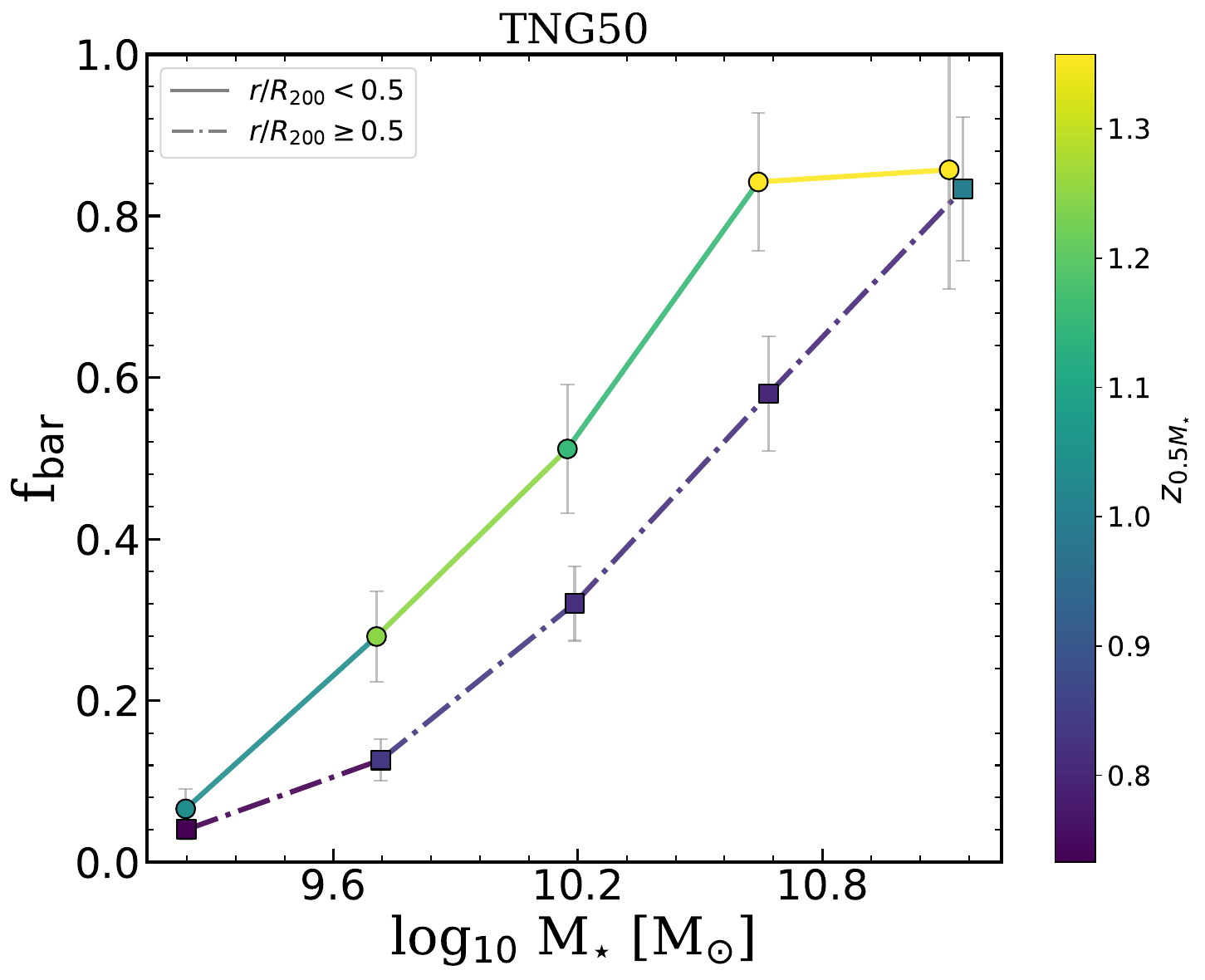}
    \subcaption{}
    \label{fbar_colormapTNG50}
\end{subfigure}%
\hspace{-0.01\linewidth}
\begin{subfigure}{0.5\linewidth}
    \centering
    \includegraphics[width=\linewidth]{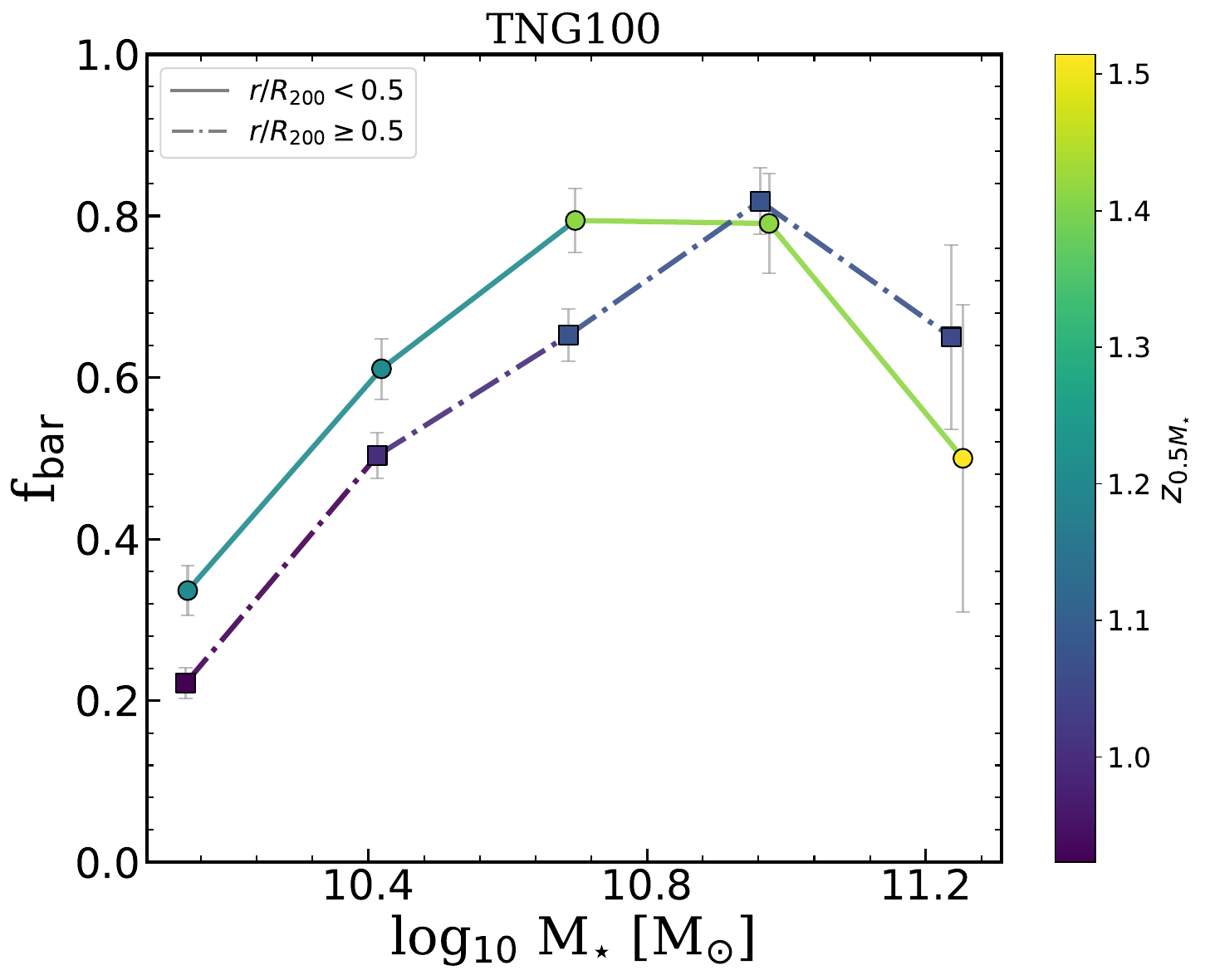}
    \subcaption{}
    \label{fbar_colormapTNG100}
\end{subfigure}

\caption{Bar fraction as a function of stellar mass for TNG50 (left) and TNG100 (right). In both panels, we distinguish galaxies located within and beyond 0.5 times the virial radius (solid and dotted lines, respectively). In both cases, the bar fraction is higher for galaxies near the cluster centre, which also assembled their stars earlier than those in the outskirts.} 
\label{fbar_colormap}
\end{figure*}

In Fig. \ref{fbar_colormap}, we extend this analysis by showing the bar fraction as a function of stellar mass for TNG50 (left) and TNG100 (right). We adopt stellar mass as the primary variable given that it is the strongest intrinsic property correlated with bar fraction, and because the stellar mass assembly time is itself closely linked to the final galaxy mass. In each simulation, galaxies are divided according to their position within or beyond half the virial radius (solid and dotted lines, respectively), while the colour map indicates the stellar mass assembly redshift, $z_{0.5 \mathrm{M}_{\star}}$.

As expected from the trends discussed above, galaxies located at larger cluster-centric distances tend to have assembled later than those near the core. However, the figure reveals that the probability of hosting a bar is primarily driven by stellar mass and assembly history: at fixed stellar mass, galaxies with earlier assembly times exhibit higher bar fractions. The apparent environmental dependence is therefore likely secondary, reflecting the fact that more massive galaxies with earlier assembly histories are preferentially found in the inner regions of clusters. In this context, the observed trends suggest that bar formation is mainly regulated by internal evolutionary processes linked to galaxy mass and assembly, while environmental effects may play a more indirect role (e.g. \citealt{lokas2016tidally, cervantes2013galactic, rosas2020buildup}).

\section{Summary and conclusions}
\label{discussion}
In this work, we study the presence of bars in the TNG100 and TNG50 simulations of the IllustrisTNG project at $z=0$. Specifically, we analyze late-type, rotation-dominated galaxies with $k_{\mathrm{rot}} \geq 0.4$ and stellar masses above $10^9$ M$_{\odot}$ in TNG50 and $10^{10}$ M$_{\odot}$ in TNG100. Our samples, which satisfy these two constraints, contain $2639$ galaxies in TNG50 and $4953$ in TNG100. To identify bar structures, we use the Fourier decomposition method. To analyze the morphological components of galaxies and the dependence on the bar presence, we use the \texttt{MORDOR} algorithm (\citealt{zana2022mordor}) that splits the stellar particles according to their energy and circularity. For the analysis, we focus on different environments defined by halo mass: low-mass (M$_h \leq 10^{13}$ M$_{\odot}$), intermediate-mass ($10^{13} < $ M$_h < 10^{14}$ M$_{\odot}$), and high-mass (M$_h \geq 10^{14}$ M$_{\odot}$) haloes. We then examine bar presence and bar properties as a function of cluster-centric distance, $r/R_{200}$, for the satellite galaxies in our samples. Restricting the analysis to satellites with cluster-centric distance lower than $2$ R$_{200}$, the final samples are reduced to $859$ galaxies in TNG50 and $1,719$ in TNG100. The number of haloes in each mass range is provided in Table \ref{numhaloes}. We also compare the stellar-mass assembly histories of barred and unbarred galaxies. Our main results are summarized as follows.

\begin{enumerate}
    \item We reproduce the morphology-density relation for the satellite galaxies in our subsamples as observed in Fig. \ref{ThinDTNG50y100}, where galaxies become more disky towards the outskirts of the clusters in both simulations. Also, we find that the bar fraction slightly increases with the thin disc component only at lower cluster-centric distances ($r/R{200} \lesssim 0.2$), while at intermediate and larger values the trend flattens or even decreases for massive haloes in TNG100.
    
    \item Bar fraction increases with stellar mass (Fig. \ref{fbar_logM_50_100}) in both simulations and for both central and satellite galaxies. In TNG50, there is no noticeable difference in the bar fraction behaviour between centrals and satellites. However, in TNG100 a difference appears, with central galaxies exhibiting a lower bar fraction at fixed stellar mass than satellites.
    \item When analysing the bar fraction and its properties as a function of cluster-centric distance, we find that bar presence slightly decreases towards the outskirts of the clusters, suggesting that bars are more common in denser environments located in the inner regions of clusters. In addition, at fixed cluster-centric distance, more massive haloes also exhibit higher bar fractions. This trend is more evident in TNG100 and for low- and intermediate-mass haloes in TNG50. Regarding the normalized bar size, we find that it decreases with cluster-centric distance in both simulations. In TNG100, this trend is observed in all halo mass ranges, while in TNG50 it is observed only for the massive and intermediate halo mass ranges. In TNG100, bars show a mild tendency to be stronger at smaller $r/R_{200}$; no clear trend is evident in TNG50.
    \item The bar fraction shows only a weak dependence on environment or cluster-centric distance, and can be largely understood in terms of its stronger correlations with stellar mass and assembly history. Galaxies that are more massive and assembled earlier are more likely to host bars, and these systems are preferentially found both in the inner regions of clusters and in more massive haloes (Fig. \ref{z05logMs_vs_rR200}). As a result, the observed environmental trends in bar fraction are likely indirect, reflecting the underlying dependence on intrinsic galaxy properties rather than a primary environmental effect.
    
\end{enumerate}

Our results shed light on the question of how bars are affected by the environment in which galaxies reside. Although we did not find a strong dependence of bar presence on the position of galaxies within the cluster, we do find differences when characterising them by the mass of their host halo. In addition, we find that the time at which galaxies assembled their stellar component has a stronger effect on whether or not they host a bar. More massive galaxies, which are typically located in the inner regions of the cluster and in more massive clusters, may have experienced disc instabilities that triggered the formation of a stellar bar. Galaxies that assembled their stellar component earlier may also have undergone such instabilities. In future work,  we will investigate the evolutionary pathways that give rise to the trends observed at $z=0$, with the aim of identifying the physical mechanisms that drive bar formation and morphological transformations across different environments. This approach is motivated by recent studies (\citealt{lokas2026diverse}) highlighting the diversity of bar formation channels in dense environments, suggesting that multiple evolutionary routes can lead to similar present-day bar properties.

\section*{Acknowledgements}

We thank the referee for raising insightful queries that have led to an enhancement of the manuscript. Karol Chim-Ramirez and Bernardo Cervantes Sodi acknowledge the financial support provided by PAPIIT projects IN108323 and IN111825 from DGAPA-UNAM. Karol Chim-Ramirez also acknowledges the support of the SECIHTI scholarship. Yetli Rosas Guevara acknowledges support from the Plan Propio of the University of Córdoba and the support from the Spanish Ministry of Science, Innovation and Universities (MICIU) and the Agencia Estatal de Investigación (AEI) through grant PID2025-171262NB-I00

\section*{Data Availability}

The data using in this work come from the IllustrisTNG simulations that are available in https://www.tng-project.org (\citealt{nelson2019illustristng}).


\bibliographystyle{mnras}
\bibliography{zexample} 

\appendix
\section{Bar radius dependence with distance. }
\label{AppendixA}
\begin{figure*}
\captionsetup[subfigure]{labelformat=empty}
\centering
    \begin{subfigure}{0.47\linewidth}
        \includegraphics[width=\linewidth]{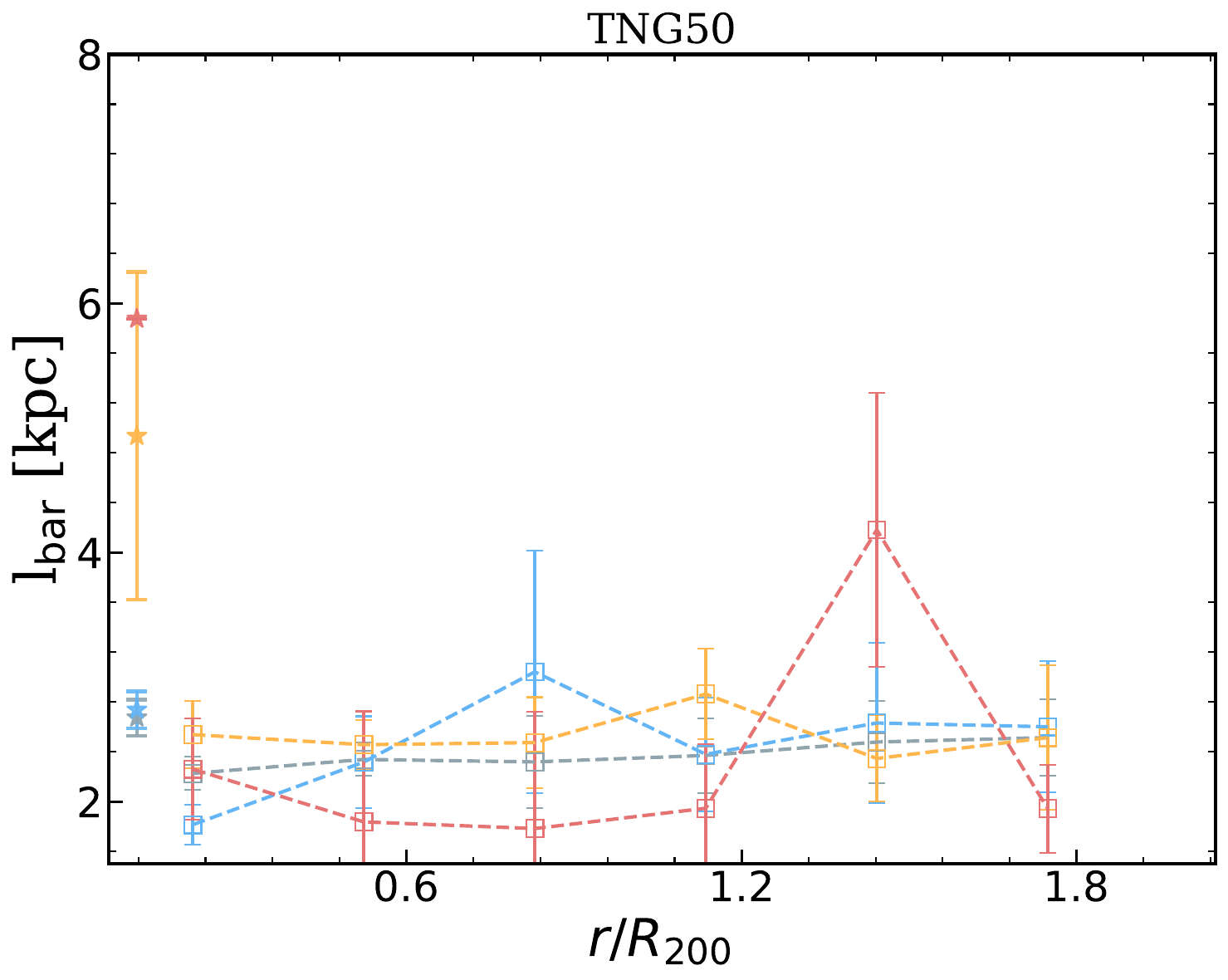}
        \subcaption{ }
        \label{fig:sfig3}
    \end{subfigure}
\hspace{-0.01\linewidth}
    \begin{subfigure}{0.47\linewidth}
        \includegraphics[width=\linewidth]{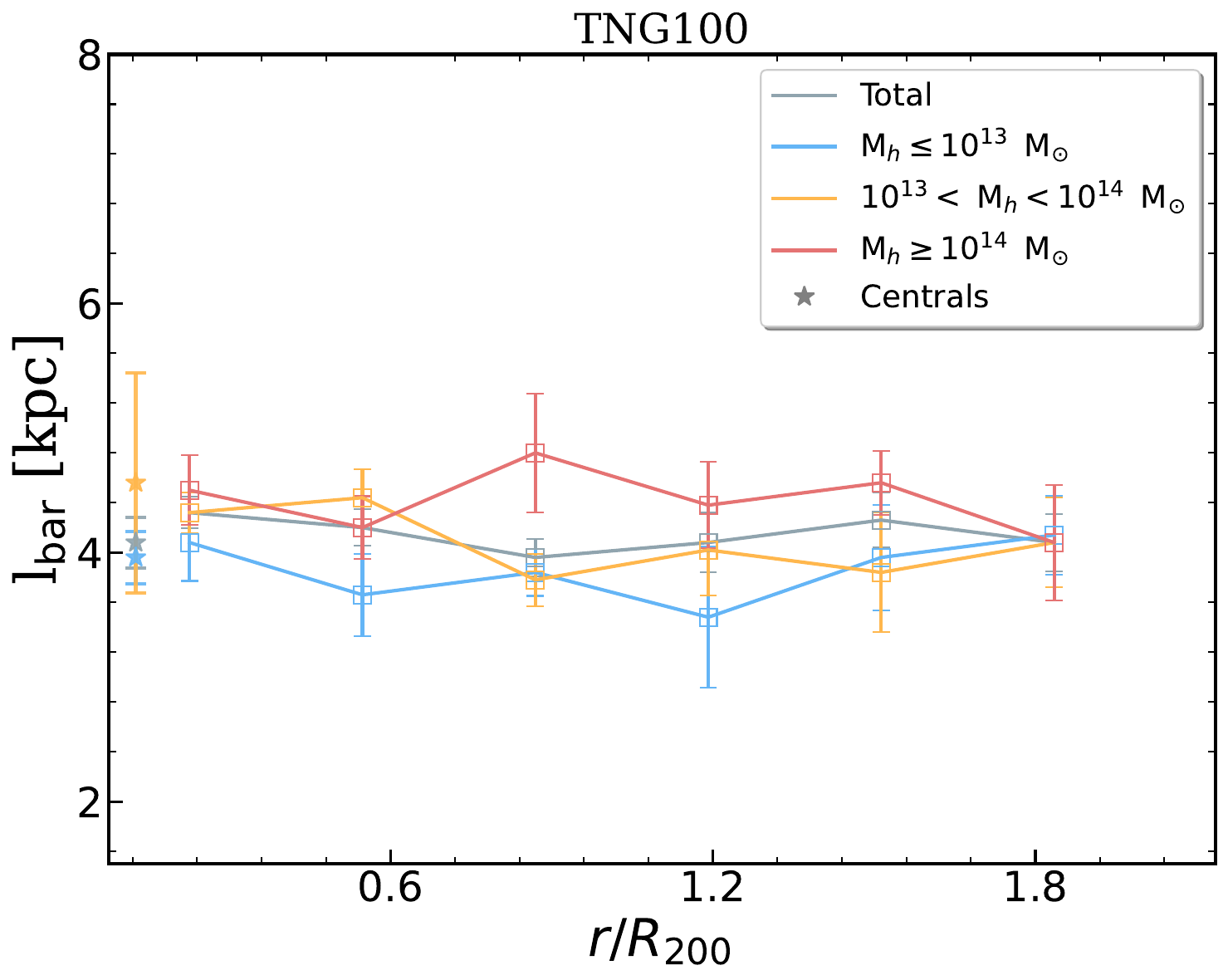}
        \subcaption{ }
        \label{fig:sfig4}
    \end{subfigure}
    \vspace{-0.2cm}
\caption{Bar radii as a function of cluster-centric distance for different halo masses. As we observe, bar sizes remain almost constant across all distance ranges in both simulations. Star markers denote central galaxies.} 
\label{lbar_cd}
\end{figure*}

To complement the bar properties as a function of cluster-centric distance presented in Fig. \ref{bar_prop}, we include the bar radius l$_{\rm bar}$ as a function of cluster-centric distance in Fig. \ref{lbar_cd}. As we observe, bar sizes remain approximately constant across all distance ranges in both simulations. Therefore, the trend observed in the central panels of Fig. \ref{bar_prop} for the normalized bar size is driven by variations in the effective radius of galaxies rather than by changes in the bar radius.

\bsp	
\label{lastpage}
\end{document}